\documentclass[journal]{IEEEtran}
\usepackage{cite}
\usepackage{xspace}
\usepackage{caption}
\usepackage{subcaption}
\usepackage{xcolor}
\usepackage[cmex10]{amsmath}
\usepackage{mathtools,amssymb,lipsum}
\usepackage{lettrine}
\usepackage{cite}
\AtBeginDocument{
  \setlength\abovedisplayskip{2pt}
  \setlength\belowdisplayskip{2pt}}
\usepackage{bm}
\usepackage{multicol, blindtext}

\usepackage{algorithm,algorithmic}
\makeatletter
\newcommand{\multilines}[1]{%
	\begin{tabularx}{\dimexpr\linewidth-\ALG@thistlm}[t]{@{}X@{}}
		#1
	\end{tabularx}
}
\makeatother
\usepackage{array}
\usepackage{url}  
\usepackage{amsthm}
\usepackage{multirow}
\usepackage{color}
\usepackage{epstopdf}
\usepackage{endnotes}
\usepackage{framed}
\usepackage{cool} 
\usepackage{enumerate} 
\usepackage{url}
\usepackage[subnum]{cases}
\usepackage[nocomma]{optidef}
\usepackage{etoolbox}

\ifCLASSINFOpdf
\else
\fi
\author{Yu~Min~Park,~Sheikh~Salman~Hassan,~\IEEEmembership{Student~Member,~IEEE},
Yan~Kyaw~Tun,~\IEEEmembership{Member,~IEEE},~Zhu~Han,~\IEEEmembership{Fellow,~IEEE,} and~Choong~Seon~Hong,~\IEEEmembership{Senior~Member,~IEEE}
\thanks{Yu Min Park, Sheikh Salman Hassan, and Choong Seon Hong  are with the Department of Computer Science and Engineering, Kyung Hee University,  Yongin-si, Gyeonggi-do 17104, Rep. of Korea, e-mails:{\{yumin0906, salman0335, cshong\}@khu.ac.kr}.}
\thanks{Yan Kyaw Tun is with Teletraffic Systems, Division of Network and Systems Engineering, School of Electrical Engineering and Computer Science, KTH Royal Institute of Technology, Brinellvägen 8, 114 28 Stockholm, Sweden, and also with the Department of Computer Science and Engineering, Kyung Hee University,  Yongin-si, Gyeonggi-do 17104, Rep. of Korea, e-mail:{\{yktun\}@kth.se}.}
\thanks{Zhu Han is with the Electrical and Computer Engineering Department, University of Houston, Houston, TX 77004, and also with the Department of Computer Science and Engineering, Kyung Hee University, Yongin-si, Gyeonggi-do 17104,  Rep. of Korea, email:{\{zhan2\}@uh.edu}.}}
\begin{document}
\title{
\LARGE Joint Trajectory and Resource Optimization of MEC-Assisted UAVs in Sub-THz Networks: A Resources-based Multi-Agent Proximal Policy Optimization DRL with Attention Mechanism} 
\maketitle
\begin{abstract}
Terahertz (THz) band communication technology will be used in the sixth-generation (6G) networks to enable high-speed and high-capacity data service demands. However, THz-communication losses arise owing to limitations, i.e., molecular absorption, rain attenuation, and coverage range. Furthermore, to maintain steady THz-communications and overcome coverage distances in rural and suburban regions, the required number of base stations (BSs) is very high. Consequently, a new communication platform that enables aerial communication services is required. Furthermore, the airborne platform supports line-of-sight (LoS) communications rather than non-LoS (NLoS) communications, which helps overcome these losses. Therefore, in this work, we investigate the deployment and resource optimization for multi-access edge computing (MEC)-enabled unmanned aerial vehicles (UAVs), which can provide THz-based communications in remote regions. To this end, we formulate an optimization problem to minimize the sum of the energy consumption of both MEC-UAV and mobile users (MUs) and the delay incurred by MUs under the given task information. The formulated problem is a mixed-integer nonlinear programming (MINLP) problem, which is NP-hard. We decompose the main problem into two subproblems to address the formulated problem. We solve the first subproblem with a standard optimization solver, i.e., CVXPY, due to its convex nature. To solve the second subproblem, we design a resources-based multi-agent proximal policy optimization (RMAPPO) deep reinforcement learning (DRL) algorithm with an attention mechanism. The considered attention mechanism is utilized for encoding a diverse number of observations. This is designed by the network coordinator to provide a differentiated fit reward to each agent in the network. The simulation results show that the proposed algorithm outperforms the benchmark and yields a network utility which is $2.22\%$, $15.55\%$, and $17.77\%$ more than the benchmarks.
\end{abstract}

\begin{IEEEkeywords}
Unmanned aerial vehicles (UAVs), mobile-edge computing, resource allocation, sub-terahertz communication, multi-agent proximal policy optimization, attention mechanism. 
\end{IEEEkeywords}
\section{Introduction}
To provide reliable network services at the target location, unmanned aerial vehicles (UAVs) are considered suitable candidates due to their efficient on-point deployment \cite{Salman_globecom}. In addition, UAVs can offer improved energy efficiency (EE), better network coverage, and increased network capacity by complementing existing terrestrial base stations \cite{own_NOMS, Salman_Iot}. However, using UAVs as communication platforms has several obstacles concerning usage, trajectory development, and network bandwidth allocation for terahertz (THz)-band full-duplex wireless communications. Because today's generation creates, transmits, and utilizes information, wireless data flow has increased dramatically in recent years \cite{traffic_forecast}. Furthermore, mobile data traffic is predicted to exceed $56$ exabytes each month, with video traffic increasing thrice \cite{traffic_forecast}. This requires the use of a wireless network with up to a terabit per second (Tbps) per device per throughput. Due to the remarkable increase in wireless communication, the exploration of a new radio spectrum has become necessary to satisfy customers' increased requirements  \cite{THz_survey} \cite{technologies7020043}.

The THz-band offers high bandwidth and data throughput compared to regular radio frequency (RF) communication bands. Currently, a new study avenue for telecom researchers and policymakers is the $0.1-10~$THz frequency range \cite{THz_pathloss}. The THz frequency range potentially offers a large bandwidth up to THz, which results in a hypothetical Tbps capacity \cite{THz_system}. Consequently, the supplied bandwidth is larger than a millimeter-wave (mmWave) system by one order of magnitude \cite{mimo_antenna}. Furthermore, THz transmissions have greater connection directionality and lower eavesdropping possibilities than mmWave signals. THz band research suggests that these frequencies have several advantages over optical frequencies, i.e., THz frequencies are excellent options for uplink communication. THz frequencies enable the propagation of non-line-of-sight (NLOS) signals to substitute reliable ones under adverse weather circumstances such as rain, fog, turbulence, and dust. In addition, an ambient sound emission from optical sources will not impact the THz frequency range and is not related to any safety or health limitations \cite{THz_link_budget}. However, despite high propagation loss, THz-band transmissions always have a huge bandwidth advantage \cite{power_allocation_THz}. This huge loss is caused by the passage of the electromagnetic (EM) signal through the medium, as well as the absorptive loss caused by the molecular absorption of atmospheric water vapor molecules \cite{THz_various_altitudes}, \cite{THz_absorbtion}.

Most prior works neglected the full-duplex communication in UAV-assisted networks at THz-band. However, there are just a few articles \cite{mobility_UAV_effect, errors_THz_UAV_positioning, Salman_ICC} about UAV communication via the THz channel. In \cite{mobility_UAV_effect}, the effect of mobility with uncertainty on THz frequency communications among flying-type UAVs was investigated. Furthermore, outside tests were conducted to assess the mobility with an uncertainty of flying UAVs of various sizes. Faced with mobile uncertainty, the possible capacity of THz connections was examined. The performance of UAV-enabled integrated access and backhaul networks within the same band was to be improved by assisting the interference control approach based on UAVs \cite{errors_THz_UAV_positioning}. Moreover, mobile user base station (BS) interactions and downlink power allocations were optimized using the fixed-point approach and particle swarm optimization (PSO). The position and orientation error constraints were used \cite{errors_THz_UAV_positioning} to study the position and orientation estimation potentials of a multiple input multiple output (MIMO) orthogonal frequency division multiple access (OFDMA) link between two THz-enabled UAVs. The performance of incorporated access and backhaul networks was improved in \cite{errors_THz_UAV_positioning} with the use of a UAV-based interference control strategy. In \cite{Salman_ICC}, the authors considered the throughput and trajectory maximization of UAVs, but they only considered UAV communication. The full-duplex communication and computation resources allocation based on multi-access edge computing (MEC)-assisted UAVs are missing.

Little prior research on optimizing UAV trajectory, spectrum, power, and energy with full-duplex communication over THz-channel has been conducted. In addition, a few key challenges in the proposed networks are as follows: how can MEC-UAVs movement patterns be evenly designed? And how should varied MEC-UAVs movement patterns and changing networks resources demand to be considered together in terms of resource allocation for the THz-communication link? To fill this research gap and answer those questions, we jointly consider optimizing these factors in our work. The key contributions of this work are summarized as follows:
\begin{itemize}
    \item We present a MEC-assisted UAVs network architecture for its data-driven task offloading to the MEC-UAVs over THz-band communication. Moreover, we investigate the effective joint trajectory optimization and resource allocation problem of MEC-UAVs in sub-THz networks. 
    \item With each network resource acting as an agent, we model this problem as a  cooperative multi-agent reinforcement learning (RL) mechanism. 
    \item On the one hand, resource allocation must satisfy mobile user constraints and improve quality of service (QoS); on the other hand, the trajectory of the MEC-UAVs must be optimized to the greatest extent possible. The ultimate objective is to maximize network utility (i.e., minimize network energy consumption and delay) while striking a balance between these two goals.
    \item To handle the resource allocation and trajectory difficulties in MEC-UAV networks, we present a resources-based multi-agent proximal policy optimization (RMAPPO) deep reinforcement learning (DRL) technique based on the actor-critic mechanism. 
    \item To construct distributed RMAPPO models, we employ a centralized training and distributed execution (CTDE) scheme, where each agent may monitor other agents' state and action information and then collaborate to fulfill the resource allocation job.
    \item Furthermore, the attention mechanism is introduced with RMAPPO-DRL, which enables each agent to focus on more relevant information and explicitly optimize itself.
    \item The effectiveness of the proposed approach is validated by comprehensive simulation results obtained by numerous experiments, which can maximize resource allocation and optimize the MEC-UAVs' trajectory with extremely constrained resources.
\end{itemize}

The rest of this paper is organized as follows. In Section \ref{related_work}, we present the related works of this paper. Then, the proposed system model and problem formulation are presented in Section \ref{system_model_sec}. After that, in Section \ref{solution_approach}, the proposed algorithm is described. The implementation and simulation results are described in Section \ref{simulation_results}, and finally the conclusions are presented in Section \ref{conclusion}. The main notations are summarized in Table. \ref{tab:table1}.

\section{Related Work}
\label{related_work}
\subsection{Full-Duplex Communication}
In this subsection, we provide a literature review on full-duplex communication \cite{full-duplex1,full-duplex2,Full-duplex3,Full-duplex4,Full-duplex5,Full-duplex6} . In \cite{full-duplex1}, authors investigated how to facilitate the efficient coexistence between a full-duplex access point (AP) and half-duplex mobile users in a wireless local area network. The interaction between full-duplex AP and half-duplex mobile users is investigated using an asymmetrical duplex (A-Duplex) AP as a media access control (MAC) protocol. The authors of \cite{full-duplex2} conducted a comprehensive examination of an analogous full-duplex amplify-and-forward relaying performance and suggested a channel based on power line communication (PLC) features. This channel is an analog adaptive circuit to enhance signal amplification at the relay and prevent any instability brought on by the positive feedback chain. Authors suggested a full-duplex wireless communication network in \cite{Full-duplex3} that may simultaneously support wireless power transmission from a hybrid AP in the downlink and a time-division multiple access (TDMA) scheme for mobile user devices in the uplink. 

In a cell-free massive multiple-input multiple-output (MIMO) network with incomplete channel state information (CSI) under spatially correlated channels, the authors in \cite{Full-duplex4} investigated the spectral efficiency (SE) of network-aided full-duplex communications. Large-dimensional random matrix theory is used to offer the deterministic counterparts used in the uplink total data rate with a least mean square error (MSE) receiver, the downlink total data rate with zero-forcing, and similarly the regularized zero-forcing beamforming. To jointly combine fog computing with cloud computing, the authors in \cite{Full-duplex5} develop in-band full-duplex communications. The authors build an M/M/1 queuing model to represent the computing delay by considering the statistical fluctuation of the computation time induced by co-located and concurrent workloads. In \cite{Full-duplex6}, the performance of the suggested full-duplex D$2$D-aided mmWave MIMO-NOMA with randomly scattered mobile users was investigated and evaluated under reasonable assumptions. The authors concluded that maximizing the full-duplex connection's transmission power is crucial since doing so ensures less self-interference and raises system capacity. However, none of the earlier work has discussed full-duplex communication (i.e., uplink and downlink) for the following networks that are possible by THz.

\subsection{THz-Band Communication}
In this subsection, we provide a literature review on THz-band communication \cite{THz-band1,THz-band2,THz-band3,THz-band4,mobility_UAV_effect}. The authors in \cite{THz-band1} devised a combined optimum resource allocation mechanism within the THz frequency band for self-powered devices (e.g., nano-devices) to enable energy harvesting (EH) that is based on nano-networks to achieve the largest channel capacity. The EH models and wireless communication within the THz frequency band were created utilizing the modulation technique known as time-spread on-off keying (TS-OOK). In \cite{THz-band2}, the authors use a combination of measurement, modeling, and analytical methodologies to investigate two common deployments for the interference of side lanes, composed of urban road and highway contexts, for low terahertz bands and millimeter-wave. On the side lanes, both direct interference and multipath interference of transmitting cars are taken into account. The authors in \cite{THz-band3} reported the results of a $2\times2$ MIMO channel-based THz-band line-of-sight (LoS) communication. The network infrastructure is built on the sub-harmonic mixer, which converts the vector network analyzer's measurement frequency into a range of $298$ to $313$ GHz.

The authors in \cite{THz-band4} represent a series of THz band experimental results based on propagation channel data (i.e., outdoor environment) with double-directional. The tests are carried out in one of the recently authorized frequency zones for the THz study by the Federal Communication Commission (FCC), which is between $141$ and $148.5$ GHz. The authors employed dual directional channel sounds with a frequency spectrum sounding method relying on RF-over-Fiber extensions for measurements over $100$ m in urban settings. To understand how mobility uncertainties impact mmWave or THz-band communications between airborne UAVs, the researchers in \cite{mobility_UAV_effect} took a preliminary approach. They started by conducting several field tests to ascertain the movement uncertainty of airborne UAVs of various sizes, including micro, small, and big. Next, the impact of mobility uncertainty is considered when evaluating the capability of mmWave or THz links. However, none of the earlier work has discussed THz-based full-duplex (i.e., uplink and downlink) communication based on UAVs.

\subsection{UAV-Assisted Wireless Networks}
In this subsection, we will discuss the UAV-based wireless networks in the literature \cite{UAV_1,UAV_2,UAV_3,UAV_4,UAV_5}. To make the most use of network resources, UAVs have lately gained popularity in industries including the Internet of Things (IoT), sensor networks, and three-dimensional (3D) wireless networks. In wireless sensor networks (WSNs), where energy utilization of sensors in data transmission is essential, data gathering by UAV is rather significant. The authors of  \cite{UAV_1} suggested a UAV-enabled WSN to solve this issue, in which a UAV is dispatched to collect data from networked sensors. The authors of \cite{UAV_2} proposed a wireless-powered communication network outfitted with rotary-wing UAVs to enable simultaneous energy collection and information relay to several ground mobile users. 

The authors in \cite{UAV_3} have investigated how UAV wireless mesh networks and ground-based networks share the same spectrum to increase UAV network capacity. For rotary-wing UAV-enabled full-duplex wireless-powered IoT networks, the authors in \cite{UAV_4} construct three optimization problems: a sum-throughput maximization (STM) problem, a total-time minimization (TTM) problem, and a total-energy minimizing (TEM) problem. However, none of the earlier work has discussed THz-based full-duplex (i.e., uplink and downlink) communication along with UAV trajectory optimization. The author presented a framework for a collaborative multi-UAV-assisted MEC system coupled with a MEC-enabled terrestrial base station in \cite{UAV_5}. However, real-time MEC-UAV trajectory and deployment optimization, which is critical in this network situation, is missing.

\subsection{Multi-Agent Reinforcement Learning}
In this subsection, we will discuss the multi-agent reinforcement learning (MARL) in the literature \cite{pmlr-v139-liu21j, 9509579, liu2021cooperative}. Reinforcement learning (RL) is one of the most advanced machine learning approaches, in which agents develop themselves by interacting with their environment, continually exploring, and gathering experiences to optimize their rewards \cite{pmlr-v139-liu21j}. RL has achieved significant advances in various sectors in recent years, including robots, autonomous cars, gaming, energy management, and others \cite{9509579}. It has been demonstrated to be a successful solution to resource allocation issues. In real-world situations, agents regularly collaborate to achieve the same goal. Single-agent RL techniques might fail or perform sub-optimally in these environments for various reasons, including partial observability in multi-agent systems, which is exacerbated by increasing the number of agents. MARL claims to tackle these issues with its decentralized execution and centralized training paradigm. In this paradigm, agents make judgments based on local observations, but training entails utilizing all publicly accessible knowledge.

One widely held belief in the MARL literature is that we will only train a limited number of agents, which is incorrect for many real-world MARL applications. Agents in a cooperative video game, for example, may ``generate" (i.e., be generated) or ``dead" throughout a single episode (i.e., end before the other agents). For example, a group of robots may run out of battery power, forcing one to terminate its journey before the other. In general, an agent can terminate prematurely, implying that it ceases to affect the environment or other agents in the middle of an episode. Furthermore, extra agents can be recruited during an episode. 

Existing algorithms often address these scenarios by putting inactive agents in absorbing states. Regardless of action choice, an agent stays absorbing until the entire collection of agents reaches a termination condition. Absorbing states allow current methods to teach cooperative agents to perform tasks with early termination while simplifying environment and multi-agent API implementations \cite{liu2021cooperative}. Furthermore, absorbing states allow decentralized partially observable Markov decision processes (POMDPs) and Markov games to depict terminated jobs early without modification. However, the research is currently lacking on the implications of PPO-based MADRL for UAV networks, resource allocation, and trajectory optimization \cite{MADRL_UAV}.

\begin{figure*}[t]
    \centering
    \includegraphics[width=\textwidth]{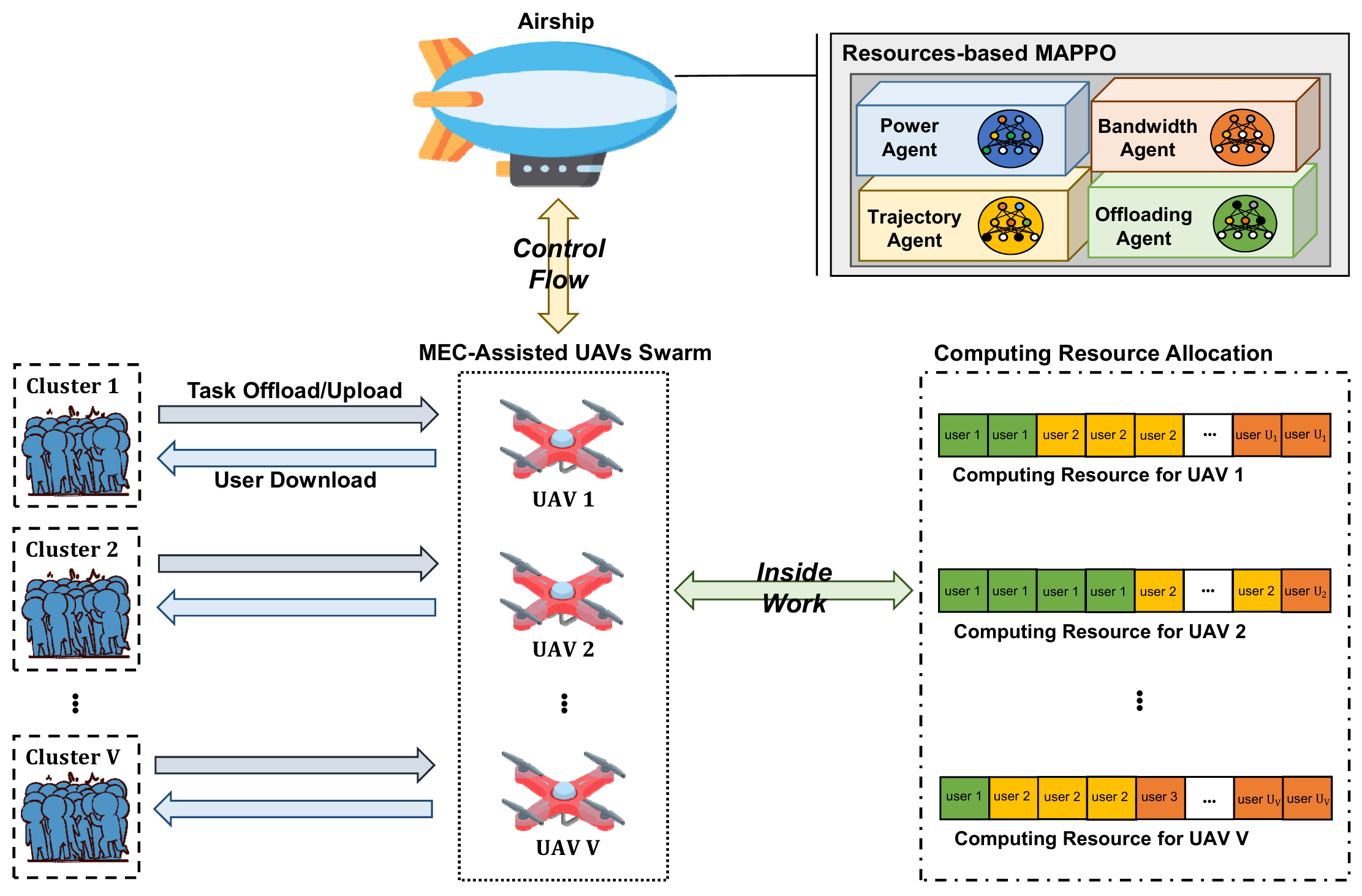}
    \caption{Illustration of joint trajectory and resource optimization of MEC-enabled UAVs in sub-THz networks.}
    \label{system_model}
\end{figure*}
\section{System Model}
\label{system_model_sec}
\subsection{Network Model}
We consider a full-duplex communication and computation system with MEC-assisted multi-UAV networks, which consists of a set $\mathcal{V}$ of $V$ UAVs attached to MEC servers and a set $\mathcal{U}$ of $U$ mobile users (MU) as presented in Fig. \ref{system_model}. We assume that MEC-UAVs are operating in the THz frequency band and the available system bandwidth is orthogonally divided amongst multiple MEC-UAVs. Furthermore, we assume that the MEC-UAVs swarm is managed by an airship\footnote{It is appropriate to deploy an airship to administer a network of MEC-UAVs since its endurance period is quite long and height is best suited for aerial control.} deployed by the network operator. To represent the dynamic nature of the network nodes, i.e., MEC-UAVs and MUs, we study the network operation within a set $\mathcal{N}$ of $N$ time slots. The network configuration is deemed fixed due to the brief duration of each time slot $n$. 

Each MEC-UAV $v$ moves its location to provide optimal wireless communication services to associated MUs while optimizing resources. The locations of MEC-UAV $v \in \mathcal{V}$ and MU $u$ are ${l}_{v}^{n}=\left[{x}_{v}^{n},{y}_{v}^{n},h_0\right]^{T}$ and ${l}_{u}^{n}=\left[{x}_{u}^{n},{y}_{u}^{n}\right]^{T}$ at the time slot $n$. Furthermore, each MU $ u \in \mathcal{U}$ has delay-sensitive computation task at each time slot $n$ as $\Psi_{u}(n)$, which can be defined as the tuple $\Psi_{u}(n) = \{D^{\textrm{pre}}_{u}(n), C^{\textrm{min}}_{u} \}~\forall u \in \mathcal{U}, ~\forall n \in \mathcal{N}$, where $D^{\textrm{pre}}_{u}(n)$ is the task's input data size and $C^{\textrm{min}}_{u}$ is the minimum CPU cycles to calculate the task data. It is difficult for MUs to compute their task locally due to the restricted computation capability of each MU and the delay limitation of the tasks. As a result, MU can transfer a portion of their tasks to MEC-UAVs over a THz-band communication link to conduct remote computing.
\begin{figure}[t]
    \centering
    \includegraphics[width=\columnwidth]{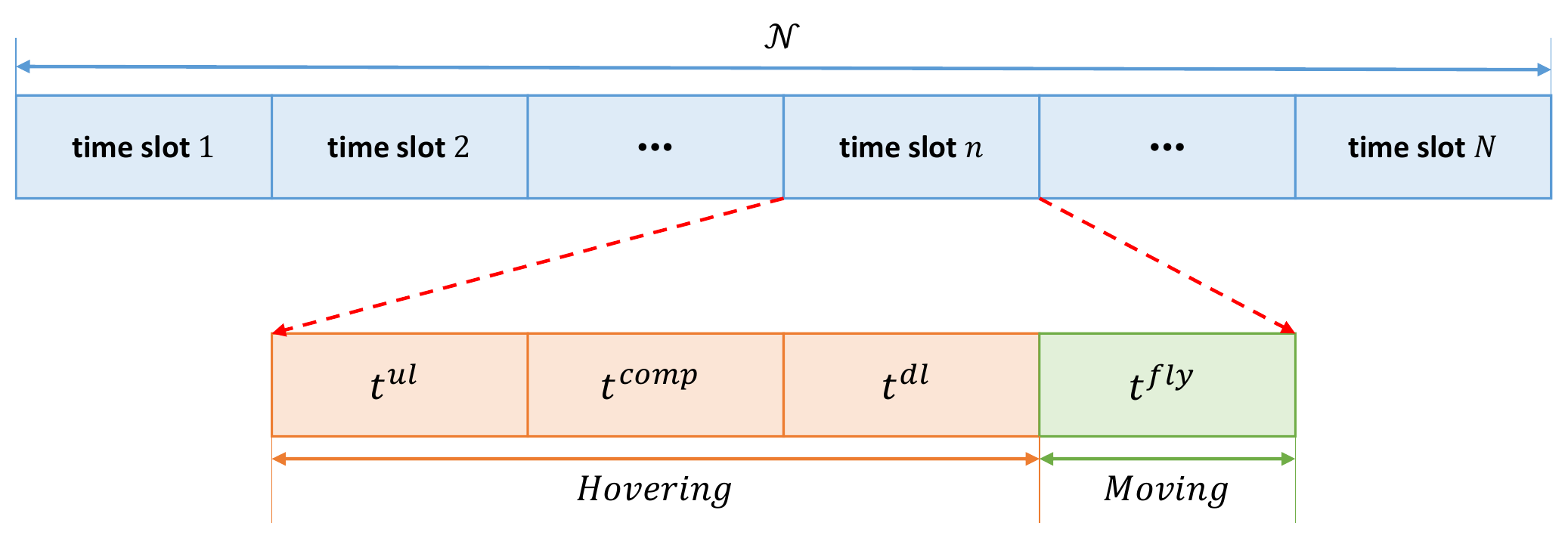}
    \caption{Structure of the network time slots.}
    \label{Optimization Framework}
\end{figure}
\begin{table}[t]
	\centering
	\caption{Summary of Notations.}
	\label{tab:table1}
	\begin{tabular}{ll}
		\hline 
		Notation & Definition\\
		\hline \hline
		$\mathcal{V}$ & Set of MEC-UAVs,  $|\mathcal{V}|= V$\\
		$\mathcal{U}$ & Set of mobile users (MUs),  $|\mathcal{U}|= U$\\
		$\mathcal{U}_v$ & Set of users of MEC-UAVs $v$ ,  $|\mathcal{U}_v|= U_v$\\
		$R^{\textrm{ul}}_{u,v}$ & Achievable data rate for the uplink transmission \\& from MU $u \in \mathcal{U}_v$ to MEC-UAVs $v \in \mathcal{V}$ \\
		$R^{\textrm{dl}}_{u,v}$ & Achievable data rate for the downlink transmission \\& from MU $u \in \mathcal{U}_v$ to MEC-UAVs $v \in \mathcal{V}$ \\
		$B^{\textrm{ul}}_v$ & Bandwidth for uplink transmission of each \\ & MEC-UAVs $v \in \mathcal{V}$ \\
		$B^{\textrm{dl}}_v$ & Total bandwidth for downlink transmission of \\& each MEC-UAVs $v \in \mathcal{V}$ \\
		$\omega_{u,v}^{\textrm{ul}}$ & Allocated subchannel for downlink transmission \\& from MEC-UAVs $v \in \mathcal{V}$ to MU $u \in \mathcal{U}_v$ \\
		$\omega_{u,v}^{\textrm{dl}}$ & Allocated subchannel for downlink transmission \\& from MEC-UAVs $v \in \mathcal{V}$ to MU $u \in \mathcal{U}_v$ \\
		$p_{0}$ & Transmit power for uplink transmission at each \\ & MU $u \in \mathcal{U}$ \\
		$p_{u,v}$ & Transmit power for downlink transmission of MU \\ & from MEC-UAVs $v \in \mathcal{V}$ to MU $u \in \mathcal{U}_v$ \\
		$t^{\textrm{ul}}_{u,v}$ & Uplink transmission delay from MEC-UAVs $v \in \mathcal{V}$ to \\& MU $u \in \mathcal{U}_v$ \\
		$t^{\textrm{dl}}_{u,v}$ & Downlink transmission delay from MEC-UAVs $v \in \mathcal{V}$ to \\ & MU $u \in \mathcal{U}_v$ \\
		$t^{\textrm{comp}}_{u,v}$ & Computation delay of MEC-UAVs $v \in \mathcal{V}$ to MU $u \in \mathcal{U}_v$ \\
		$t^{\textrm{fly}}_{v}$ & Flying time of MEC-UAVs $v \in \mathcal{V}$ \\
		$T_{u,v}$ & Total communication delay from MEC-UAVs $v \in \mathcal{V}$ to \\& MU $u \in \mathcal{U}_v$ \\
		$D^{\textrm{pre}}_{u}$ & Data size of MU $u \in \mathcal{U}$ before offloading processing \\
		$D^{\textrm{mec}}_{u}$ & Offloading data size of MU $u \in \mathcal{U}$ \\
		$D^{\textrm{in}}_{u}$ & Data size to be processed in MU $u \in \mathcal{U}$\\
		$D^{\textrm{post}}_{u,v}$ & Data size of MU $u \in \mathcal{U}_v$ after offloading processing \\& in MEC-UAVs $v \in \mathcal{V}$ \\
		$\delta^{\textrm{prog}}_{v}$ & Processing capability of MEC-UAVs $v \in \mathcal{V}$\\
		$\alpha_{u}$ & Offloading parameter of MU $u \in \mathcal{U}$'s data \\
		$E^{\textrm{up}}_{u,v}$ & Energy consumption uplink transmission of \\ &MEC-UAVs $v \in \mathcal{V}$ \\& for MU $u \in \mathcal{U}_v$ \\
		$E^{\textrm{dl}}_{u,v}$ & Energy consumption downlink transmission of \\ & MEC-UAVs $v \in \mathcal{V}$ \\& for MU $u \in \mathcal{U}_v$ \\
		$E^{\textrm{comp}}_{u,v}$ & Computing energy consumption of MEC-UAVs $v \in \mathcal{V}$ for \\& mobile users $u \in \mathcal{U}_v$ \\
		$E^{\textrm{fly}}_{v}$ & Flight energy consumption of MEC-UAVs $v \in \mathcal{V}$\\
		$E_{v}$ & Total energy consumption of MEC-UAVs $v \in \mathcal{V}$ \\
		\hline
	\end{tabular}
\end{table}
\subsection{Local Computation Model}
For remote computation, we define a decision variable as a $\alpha_u^v(n)$, which is a proportion of MU $u$'s task data that need to offload to MEC-UAV $v$ at each time slot $n$. After that the proportion of the task data that computes at MU $u$ at time slot $n$ can be defined as follows:
\begin{equation}
D^{\textrm{in}}_{u}(n) = (1-\alpha_{u}(n)) D^{\textrm{pre}}_{u}(n). \label{local_data_size}
\end{equation}
After obtaining the proportion of task data $D^{\textrm{in}}_{u}(n)$ for local computation, we can find local computation delay of MU $u$ at time slot $n$ is as follows:
\begin{equation}
t^{\textrm{comp}}_{u}(n)=  {{\beta_u D^{\textrm{in}}_u (n)}\over{c_u^{\textrm{in}}}}\label{local_comp_delay},
\end{equation}
where $\beta_u$ denotes the desired CPU cycles to compute $1$-bit of input task data and $c^{\textrm{in}}_u$ is the computation capacity (i.e., cycles/s) that MU $u$ utilizes at each time slot $n$. After obtaining the local task data $D^{\textrm{in}}_{u}(n)$ and its computation time $t^{\textrm{comp}}_{u}(n)$, we can calculate the local computation energy consumption at MU $u$ to execute the task at time slot $n$ is as follows:
\begin{equation}
E^{\textrm{comp}}_{u}(n) = q_{u} \left(c^{\textrm{in}}_{u}\right)^2 \beta_{u} D^{\textrm{in}}_u (n),
\end{equation}
where $q_u$ refers to the chip architecture constant installed in MU $u$.
\subsection{MEC-UAVs Computation Model}
In order to execute MU $u$ task data to the MEC-UAV, i.e., remote computation, each MU $u$ offloads a fraction of the computation task to MEC-UAV $v$ by using the THz-band communication link. Therefore, the achievable data rate for the uplink communication link from MU $u$ to MEC-UAV $v$ at each time slot $n$ is \cite{xu2021joint}:
\begin{equation}
R^{\textrm{ul}}_{u,v} (n) = \omega_{u,v}^{\textrm{ul}} B_v^{\textrm{ul}} \log_2 \left( 1 + {{p^{\textrm{ul}}_u h_0} \over {\omega_{u,v}^{\textrm{ul}} B_v^{\textrm{ul}} (d^{n}_{u,v})^2 e^{a d^{n}_{u,v}} \sigma^2}} \right), \label{eq1}
\end{equation}
where $\omega_u^{\textrm{ul}}$ denotes the fraction of bandwidth (i.e., THz-band) designated to MU $u$, $B^{\textrm{ul}}$ refers to the total bandwidth for the uplink transmission of MEC-UAV $v$, $p^{\textrm{ul}}_u$ is the transmit power of MU $u$, $\sigma^2$ represents the Gaussian noise power, $a$ indicates the molecular absorption constant for THz-frequency, $d_n$ is the distance between MU $u$ and MEC-UAV $u$, $h_0$ is the channel gain at a reference distance $d_0 = 1$ m, and $d^{n}_{u,v}=\sqrt{\left(h_0 \right)^{2} + {\Vert {l}_{v}^{n} - {l}_{u}^{n} \Vert}^{2}}$ is distance between MEC-UAV $v$ and MU $u$ at time slot $n$.

Thus, the uplink communication delay experienced by MU $u$ associated with MEC-UAV $v$ at time slot $n$ is as follows:
\begin{equation}
t^{\textrm{ul}}_{u,v} (n) = {{D^{\textrm{mec}}_{u} (n)} \over{R^{\textrm{ul}}_{u,v} (n)}},\label{uplink_delay}
\end{equation}
where $D^{\textrm{mec}}_{u}(n) = \alpha_{u}(n) D^{\textrm{pre}}_{u} (n)$ denotes the task data size of MU $u$ that is offloaded to MEC-UAV $v$ at time slot $n$. Then, the uplink communication energy consumption for MU $u$ at time slot $n$ can be defined as follows:
\begin{equation}
E^{\textrm{up}}_{u,v}(n) = t^{\textrm{ul}}_{u_v} (n) \cdot p^{\textrm{ul}}_u\label{ulink_energy}.
\end{equation}
Following that, MU $u$ offloaded task data (e.g., $D^{\textrm{mec}}_{u} (n)$) is executed at MEC-UAV $v$. 

As a result, the computation delay incurred by MU $u$ in order to successfully complete the computation of its offloaded task data at MEC-UAV during time slot $n$ can be calculated as follows \cite{tun2020energy}: 
\begin{equation}
t^{\textrm{comp}}_{u,v} (n)= {{\beta^v D^{\textrm{mec}}_u (n)}\over{c_{u,v}^{\textrm{mec}}}},
\end{equation}
where $\beta_v$ denotes the required CPU cycles to compute $1$-bit of input data and $c_{uv}^{\textrm{mec}}$ refers to the computation capacity of MEC-UAV $v$ that is allocated to compute the offloaded task of MU $u$. 

Additionally, the energy consumption at MEC-UAV $v$ to compute the task data $D^{\textrm{mec}}_{u}(n)$) that was offloaded from MU $u$ at time slot $n$ can be defined by following \cite{hu2018joint}:
\begin{equation}
E^{\textrm{comp}}_{u,v}(n) = q^{v} \left(c^{\textrm{mec}}_{u,v}\right)^2 \beta^v D^{\textrm{mec}}_u (n),
\end{equation}
where $q^v$ refers to the chip architecture constant that is installed at MEC-UAV $v$. 

Following computation of the offloaded task $D^{\textrm{mec}}_u (n)$ at MEC-UAV $v$, MU $u$ receives the processed result through downlink communication. Consequently, the achievable data rate for the downlink communication from MEC-UAV $v$ to MU $u$ at time slot $n$ is as follows:
\begin{equation}
R^{\textrm{dl}}_{u,v} (n) = \omega_{u,v}^{\textrm{dl}} B_v^{\textrm{dl}} \log_2 \left( 1 + {{p_{u,v} h_0} \over {\omega_{u,v}^{\textrm{dl}} B_v^{\textrm{dl}} (d^{n}_{u,v})^2 e^{a d^{n}_{u,v}} \sigma^2}} \right),\label{eq1}
\end{equation}
where $\omega_u^{v,\textrm{dl}}$ denotes the allocated THz-bandwidth from MEC-UAV $v$ to MU $u$, $B_v^{\textrm{dl}}$ refers to the total THz-bandwidth for the downlink communication of MEC-UAV $v$, and $p^{v,\textrm{dl}}_u$ represents the transmit power of MEC-UAV $v$ to MU $u$.

As a result, MU $u$ during time slot $n$ encountered the following downlink communication delay:
\begin{equation}
t^{\textrm{dl}}_{u,v} (n) =  {{D^{\textrm{post}}_{u,v} (n)}\over{R^{\textrm{dl}}_{u,v} (n)}},\label{eq1}
\end{equation}
\begin{equation}
D^{\textrm{post}}_{u,v} (n) = \delta^{\textrm{prog}}_{v} D^{\textrm{mec}}_u (n),
\end{equation}
where $\delta^{\textrm{prog}}_{v}$ denotes a constant between $0$ and $1$, and $D^{\textrm{mec}}_u (n)$ presents the data size of processed task of MU $u$ that is transmitted back from MEC-UAV $v$ at time slot $n$. Next, MEC-UAV $v$ uses the following amount of energy to communicate the processed result back to MU $u$ at time slot $n$:
\begin{equation}
E^{\textrm{dl}}_{u,v}(n) = t^{\textrm{dl}}_{u,v} (n) \cdot p_{u,v}\label{eq1}.
\end{equation}
In addition, we evaluate the flight energy consumption, which is required to keep MEC-UAV $v$ afloat and, if necessary, to continue its movement. The flight energy consumption of a MEC-UAV $v$ is determined by its speed and design. Therefore, the flight energy consumption of a rotary-wing MEC-UAV $v$ with speed $v_{v}$ at time slot $n$ is calculated as follows \cite{li2020delay}:
\begin{equation}
E^{\textrm{fly}}_{v}(n) = t^{\textrm{fly}}_{v} \cdot v_{v}(n) \left[ c_{1}(v_{v}(n))^{2}+{c_{2}\over{(v_{v}(n))^{2}}} \right],\label{eq1}
\end{equation}
where parameters $c_1$ and $c_2$ are determined by the MEC-wing UAV's area, weight, and air density. Additionally, we solely consider the MEC-UAV's flying energy and exclude its hovering energy for the sake of generality. 

In order to complete the execution of MUs' tasks at time slot $n$, the energy consumption at both MEC-UAV $v$ and its associated MUs $U_v$ can be calculated as follows:  
\begin{multline}
E_{v}(n) = E^{\textrm{fly}}_{v}(n) + \sum^{U_v}_{u=1} \Big[ E^{\textrm{comp}}_{u}(n)+ E^{\textrm{ul}}_{u,v}(n) + \\ E^{\textrm{dl}}_{u,v}(n) + E^{\textrm{comp}}_{u,v}(n) \Big].
\end{multline}
The following total delay was encountered by MU $u$ associated with MEC-UAV $v$ in order to finish the execution of its task at time slot $n$:
\begin{equation}
T_{u,v} (n)= t^{\textrm{comp}}_{u} (n) + t^{\textrm{ul}}_{u,v} (n)+ t^{\textrm{comp}}_{u,v} (n) + t^{\textrm{dl}}_{u,v} (n). \label{remote_delay}
\end{equation}
Based on the findings of the preceding study, we can now outline our energy and delay minimization problem in the next subsection.
\subsection{Problem Formulation}
In this subsection, we will define the detailed problem formulation based on the proposed system model. This work's major goal is to meet each MU's QoS requirements while minimizing the total energy consumption and network delay of both MEC-UAV and MUs, which are incurred under the tasks' supplied information (i.e., MUs' task data size, $2$D position, and UAV-MEC association). Therefore, we can define our optimization problem as follows:
\begin{mini!}[2]                 
		{\substack{\boldsymbol{\omega^{\textrm{ul}}}, \boldsymbol{\omega^{\textrm{dl}}}, \boldsymbol{p},\\ \boldsymbol{c^{\textrm{in}}}, \boldsymbol{c^{\textrm{mec}}}, \boldsymbol{L}, \boldsymbol{\alpha}}}                             
		{\sum_{n=1}^{N} \sum_{v=1}^{V} \left[ \eta E_v(n) + (1-\eta)\sum_{u=1}^{U_v}  T_{u,v}(n) \right]  }{\label{opt:P1}}{\textbf{P1:}}
		\addConstraint{\bigcup_{v=1}^V U_v = \mathcal{U},  \label{C1}}
		\addConstraint{U_i \cap U_j = \varnothing, \forall i\neq j \in \mathcal{V}, \label{C2}}
		\addConstraint{
		\begin{aligned}
		R_{u,v}^{\textrm{ul}}(&\boldsymbol{\omega^{\textrm{ul}}}) \geq R^{\textrm{min}},  \\&~\forall u \in \mathcal{U}_{v}\forall {v} \in \mathcal{V},
		\end{aligned}
		\label{C3}}
		\addConstraint{
		\begin{aligned}
		R_{u,v}^{\textrm{dl}}(&\boldsymbol{\omega^{\textrm{dl}}, p, L}) \geq R^{\textrm{dl}}_{\mathbf{min}}, \\&~\forall {u} \in \mathcal{U}_{v}, \forall {v} \in \mathcal{V},
		\label{C4}
		\end{aligned}}
		\addConstraint{ \sum_{u=1}^{U_v} \omega_{u,v}^{\textrm{ul}}  \leq 1 , \ \forall v \in \mathcal{V},   \label{C5}}
		\addConstraint{ 0 \leq \omega_{u,v}^{\textrm{ul}} \leq 1 , \ \forall u \in \mathcal{U}, \label{C6}}
		\addConstraint{ \sum_{u=1}^{U_v}  \omega_{u,v}^{\textrm{dl}} \leq  1 , \ \forall v \in \mathcal{V},  \label{C7}}
		\addConstraint{ 0 \leq \omega_{u,v}^{\textrm{dl}} \leq 1 , \ \forall u \in \mathcal{U}, \label{C8}}
		\addConstraint{\sum_{u=1}^{U_v} \omega_{u,v}^{\textrm{dl}} B_v^{\textrm{dl}}p_{u,v} \leq P_v^{\mathbf{max}}, \  \forall v \in \mathcal{V},  \label{C9}}
		\addConstraint{0 \leq p_{u,v} \leq P_v^{\mathbf{max}}, \forall u \in \mathcal{U}_{v}, \forall v \in \mathcal{V}, \label{C10}}
		\addConstraint{ \sum_{u=1}^{U_v} c^{\textrm{mec}}_{u,v} \leq C_v^{\mathbf{max}},  \  \forall v  \in \mathcal{V}, \label{C11} }
		\addConstraint{ C_u^{\textrm{min}} \leq c^{\textrm{mec}}_{u,v} \leq C_v^{\mathbf{max}},  \forall u \in \mathcal{U}_{v}, \forall v \in \mathcal{V} \label{C12}},
		\addConstraint{ C_u^{\textrm{min}} \leq c^{\textrm{in}}_u \leq C_u^{\mathbf{max}},  \forall u \in \mathcal{U}, \label{C13}}
		\addConstraint{
		\begin{aligned}
		\lVert {l}^{n}_{v} - {l}^{n}_{\Bar{v}} \rVert &\geq L^{\mathbf{min}} ,\\&~\forall v\neq \Bar{v} \in \mathcal{V} ,\forall n \in \mathcal{N},
		\end{aligned}
		\label{C14}}
		\addConstraint{
		\begin{aligned}
		{\lVert {l}^{n+1}_{v}-{l}^{n}_{v} \rVert \over{D}} &\leq V^\mathbf{max},   \\&~\forall v \in \mathcal{V},   \forall n \in \mathcal{N},
		\end{aligned}
		\label{C15}}
		\addConstraint{0 \leq \alpha_{u}(n) \leq 1, \forall u \in \mathcal{U}, \forall n \in \mathcal{N}, \label{C16}}
\end{mini!}
where the objective function defines the trade-off between task energy consumption and computation delay at each MEC-UAV, which is controlled by the parameter $\eta$. Constraints (\ref{C1}) and (\ref{C2}) guarantees that each MU can be associated with only one MEC-UAV. Constraints (\ref{C3}) and (\ref{C4}) ensure that each MU and MEC-UAV must fulfill the QoS for uplink and downlink achievable data rate, respectively. Constraints (\ref{C5}) and (\ref{C6}) are the constraints for the proportion of THz-bandwidth allocation from MU $u$ to its associated MEC-UAV $v$ for uplink communication. Similarly, constraints (\ref{C7}) and (\ref{C8}) are the constraints for the proportion of THz-bandwidth allocation from MEC-UAV $v$ to its associated MU $u$ for downlink communication. Constraints (\ref{C9}) and (\ref{C10}) are constraints for the transmit power allocation from MEC-UAV $u$ to its associated MU $u$. Moreover, constraints (\ref{C11}) and (\ref{C12}) are the constraint for the computation resources allocation from each MEC-UAV $v$ to its associated MU $u$. Similarly, constraint (\ref{C13}) is the constraint that guarantees that each MU $u$'s local computation power should not exceed the maximum computation budget. Moreover, we provide the constraint for each MEC-UAV's trajectories, i.e, constraint (\ref{C14}) represents that the distance between two MEC-UAVs should be greater than the threshold (i.e., safe and ensures energy efficiency) distance, which is defined as $L^{\mathbf{min}}$, and constraint (\ref{C15}) ensures that speed of each MEC-UAV $v$ should be less than the threshold values which ensures the QoS for MUs. Finally, the constraint (\ref{C16}) decides how much data proportion of each MU $u$ will be offloaded to MEC-UAV $v$. To solve this proposed problem, we provide a solution approach in the next section.
\section{Towards Resources-Based Multi-Agent Proximal Policy Optimization Deep Reinforcement Learning Solution Approach}
\label{solution_approach}
It can be observed that the formulated problem in (\ref{opt:P1}) is a mixed-integer nonlinear programming (MINLP) problem, which is a non-polynomial-time (NP-hard) problem. Therefore, it is not possible to get an optimal solution within polynomial time. Thus, to solve the problem realistically, we decompose the main problem into two subproblems as follows. First in Subsection \ref{4_A}, since the optimization for computation resource allocation takes place within each network device (i.e., MEC-UAVs and MUs), the subproblem for computation resource allocation is introduced. Second in Subsection \ref{4_B}, through the resources-based multi-agent proximal policy optimization (RMAPPO) deep reinforcement learning method, we address the optimization problem for communication and offloading resource allocation, and the MEC-UAVs' trajectories.
\subsection{Computation Resource Allocation for MEC-UAVs and MUs}
\label{4_A}
In this subsection, we tackle the subproblem of local (MUs-based) and offloaded (MEC-UAVs-based) computation resource allocation and need to find the optimal solution for each. The subproblem can be defined as follows: 
\begin{mini!}[2]                 
		{\substack{\boldsymbol{c^{\textrm{in}}}, \boldsymbol{c^{\textrm{mec}}}}}                             
		{\sum_{n=1}^{N} \sum_{v=1}^{V} \left[ \eta E_v(n) + (1-\eta)\sum_{u=1}^{U_v}  T_{u,v}(n) \right]  }{\label{opt:P}}{\textbf{P1.1:}}
		\addConstraint{\bigcup_{v=1}^V U_v = \mathcal{U},}
		\addConstraint{U_i \cap U_j = \varnothing, \forall i\neq j \in \mathcal{V},}
		\addConstraint{ \sum_{u=1}^{U_v} c^{\textrm{mec}}_{u,v} \leq C_v^{\mathbf{max}},  \  \forall v  \in \mathcal{V}, }
		\addConstraint{  C_u^{\textrm{min}} \leq c^{\textrm{mec}}_{u,v} \leq C_v^{\mathbf{max}},  \forall u \in \mathcal{U}_{v}, \forall v \in \mathcal{V}},
		\addConstraint{  C_u^{\textrm{min}} \leq c^{\textrm{in}}_u \leq C_u^{\mathbf{max}},  \forall u \in \mathcal{U}. }
\end{mini!}
It can be observed that the proposed subproblem is convex in nature and can be solved with any standard optimization solver. Therefore, to solve this problem, we can utilize the CVXPY toolkit \cite{diamond2016cvxpy}. After solving this problem, we can obtain the optimal computation resource allocation solutions i.e., $\boldsymbol{c}^{*\mathrm{in}}, \boldsymbol{c}^{*\mathrm{mec}}$.

\subsection{Resources-based Multi-Agent Proximal Policy Optimization Deep Reinforcement Learning for Resources Allocation and Trajectory Optimization}
\label{4_B}
In this subproblem, we will find the optimal decision variables for communication resource allocation (i.e., uplink and downlink bandwidth and transmit power), MEC-UAVs trajectories, and offloading data proportions. The subproblem can be defined as follows: 
\begin{mini!}[2]                 
		{\substack{\boldsymbol{\omega^{\textrm{ul}}}, \boldsymbol{\omega^{\textrm{dl}}}, \boldsymbol{p},\\ \boldsymbol{L}, \boldsymbol{\alpha}}}                             
		{\sum_{n=1}^{N} \sum_{v=1}^{V} \left[ \eta E_v(n) + (1-\eta)\sum_{u=1}^{U_v}  T_{u,v}(n) \right]  }{\label{opt:SP2}}{\textbf{P1.2:}}
		\addConstraint{\bigcup_{v=1}^V U_v = \mathcal{U},}
		\addConstraint{U_i \cap U_j = \varnothing, \forall i\neq j \in \mathcal{V},}
		\addConstraint{
		\begin{aligned}
		R_{u,v}^{\textrm{ul}}(&\boldsymbol{\omega^{\textrm{ul}}}) \geq R^{\textrm{ul}}_{\mathbf{min}},  \\&~\forall u \in \mathcal{U}_{v}, \forall {v} \in \mathcal{V},
		\end{aligned}
		}
		\addConstraint{
		\begin{aligned}
		R_{u,v}^{\textrm{dl}}(&\boldsymbol{\omega^{\textrm{dl}}, p, L}) \geq R^{\textrm{dl}}_{\mathbf{min}}, \\&~\forall {u} \in \mathcal{U}_{v}, \forall {v} \in \mathcal{V},
		\end{aligned}
		}
		\addConstraint{ \sum_{u=1}^{U_v} \omega_{u,v}^{\textrm{ul}}  \leq 1 , \ \forall v \in \mathcal{V},}
		\addConstraint{ 0 \leq \omega_{u,v}^{\textrm{ul}} \leq 1 , \ \forall u \in \mathcal{U},}
		\addConstraint{ \sum_{u=1}^{U_v}  \omega_{u,v}^{\textrm{dl}} \leq  1 , \ \forall v \in \mathcal{V},}
		\addConstraint{ 0 \leq \omega_{u,v}^{\textrm{dl}} \leq 1 , \ \forall u \in \mathcal{U},}
		\addConstraint{\sum_{u=1}^{U_v} \omega_{u,v}^{\textrm{dl}} B_v^{\textrm{dl}}p_{u,v} \leq P_v^{\mathbf{max}}, \  \forall v \in \mathcal{V},}
		\addConstraint{0 \leq p_{u,v} \leq P_v^{\mathbf{max}}, \forall u \in \mathcal{U}_{v}, \forall v \in \mathcal{V},}
		\addConstraint{
			\begin{aligned}
		\lVert {l}^{n}_{v} - {l}^{n}_{\Bar{v}} \rVert &\geq L^{\mathbf{min}} ,\\&~\forall v\neq \Bar{v} \in \mathcal{V} ,\forall n \in \mathcal{N},
		\end{aligned}
		\label{C14}}
		\addConstraint{
		\begin{aligned}
		{\lVert {l}^{n+1}_{v}-{l}^{n}_{v} \rVert \over{D}} &\leq V^\mathbf{max},   \\&~\forall v \in \mathcal{V},   \forall n \in \mathcal{N},
		\end{aligned}
		}
		\addConstraint{0 \leq \alpha_{u}(n) \leq 1, \forall u \in \mathcal{U}, \forall n \in \mathcal{N}.}
\end{mini!}
It can be observed that the subproblem is complex in nature. To solve this problem, we propose a low-complexity multi-agent deep reinforcement learning algorithm based on proximal policy optimization, which can be discussed in the sequel.
\begin{figure*}[t]
    \centering
    \includegraphics[width=\textwidth]{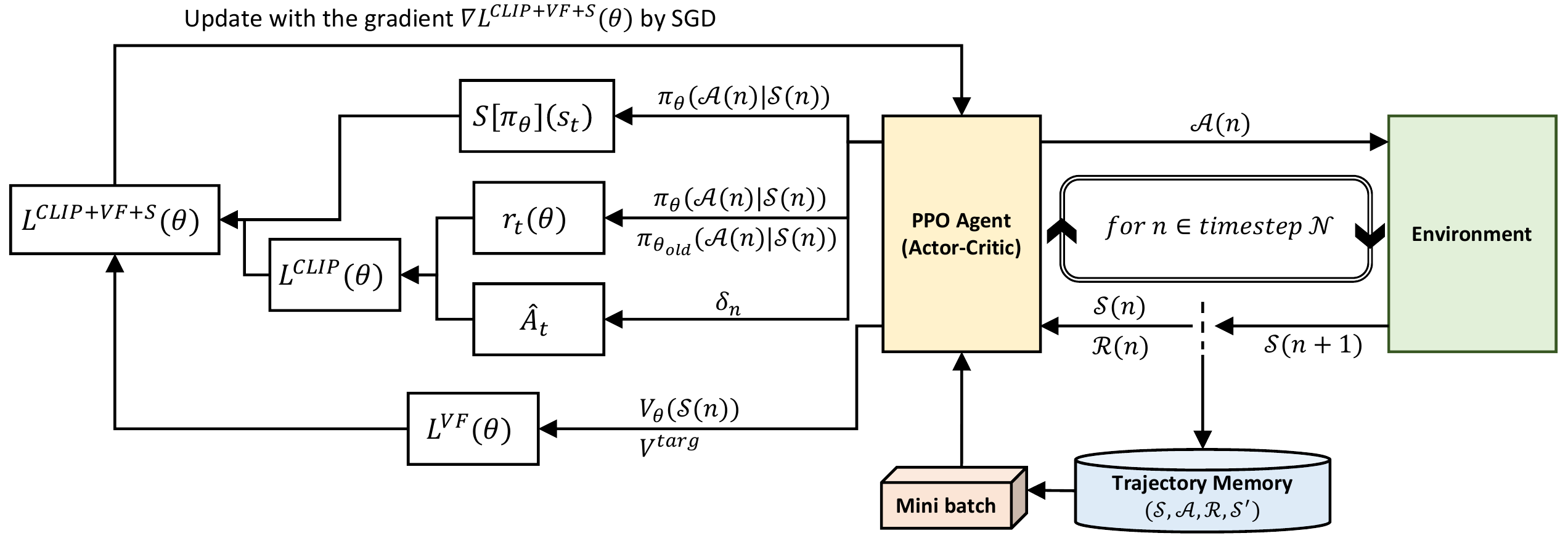}
    \caption{Learning architecture of proximal policy optimization (PPO) agent.}
    \label{structure_PPO}
\end{figure*}
\subsubsection{Proximal Policy Optimization (PPO)}
Proximal policy optimization (PPO) is a representative reinforcement learning algorithm that is simple to implement and apply in various environments and shows stable performance \cite{ppo}. PPO is a method to simplify the complex calculation of trust region policy optimization (TRPO). TRPO maximizes a surrogate objective as follows \cite{trpo}:
\begin{equation}
L^{\textrm{TRPO}}_{n}(\theta)=\hat{\mathbb{E}}_{n} \left[ {\pi_{\theta}(\mathcal{A}_{n}|\mathcal{S}_{n})\over{\pi_{\theta_{\textrm{old}}}(\mathcal{A}_{n}|\mathcal{S}_{n})}} \hat{A}_{n}  \right] = \hat{\mathbb{E}}_{n} \left[ r_{n}(\theta)\hat{A}_{n} \right],\label{L_TRPO}
\end{equation}
where $r_{n}(\theta)$ denotes the probability ratio, $\mathcal{A}_{n}$ and $\mathcal{S}_{n}$ are an action and reward in time step $n$. Without a constraint, maximization of the surrogate objective function $L^{\textrm{TRPO}}$ of TRPO would lead to an excessively large policy update. Also, the surrogate objective function $L^{\textrm{TRPO}}$ of TRPO has to find the second derivative and has a complex formula expansion \cite{trpo}. Therefore, in PPO, the shortcomings of TRPO were supplemented by performing approximately with the first derivative through the clipping method. The following is the objective function to which the clipping is applied:
\begin{equation}
L^{\textrm{CLIP}}_{n}(\theta)=\hat{\mathbb{E}}_{n} \left[ \textrm{min}(r_{n}(\theta)\hat{A}_{n}, \textrm{clip}(r_{n}(\theta), 1-\epsilon, 1+\epsilon)\hat{A}_{n}) \right],\label{L_CLIP}
\end{equation}
where $\epsilon$ is a hyperparameter and $\hat{A}_{n}$ is a truncated version of generalized advantage estimation which can be defined as follows:
\begin{equation}
    \hat{A}_{n} = \delta_{n} + (\gamma\lambda)\delta_{n+1} + \dots + (\gamma\lambda)^{N-n+1}\delta_{N-1},
\end{equation}
where $\delta_{n}=r_{n}+\gamma V(s_{n+1}) - V(s_{n})$. The function in \eqref{L_CLIP} takes a lower value when comparing the objectives used in the TRPO with the objectives to which clipping is applied. With this clipping method, we only consider the change in the probability ratio if it improves the objective. If it makes the objective worse, we leave it out.

Following that, PPO has an actor-critic network but has a network design that shares the parameters of policy and value functions. We must employ a loss function that combines the policy surrogate and an error term from the value function if we utilize a neural network design to share the parameters for the policy function and the value function. The following aim, which is improved with each iteration, is obtained by combining the policy surrogate with a value function error term:
\begin{equation}
L^{\textrm{PPO}}_{n}(\theta)=\hat{\mathbb{E}}_{n} \left[ L^{\textrm{CLIP}}_{n}(\theta) - c_{1} L^{\textrm{VF}}_{n}(\theta) + c_{2}E[\pi_\theta](s_n) \right],\label{L_PPO}
\end{equation}
where $c_1$ and $c_2$ are coefficients, $E$ denotes an entropy function, and $L^{\textrm{VF}}_{n}$ is a squared-error loss. In this objective $L^{\textrm{PPO}}$ can further be augmented by adding an entropy bonus $E$ to ensure sufficient exploration. Fig.~\ref{structure_PPO} shows the overall learning structure for the described PPO learning.

 \begin{figure*}[t]
        \begin{subfigure}[t]{0.35\textwidth}
                \captionsetup{justification=centering,singlelinecheck=false}
                \includegraphics[width=\linewidth]{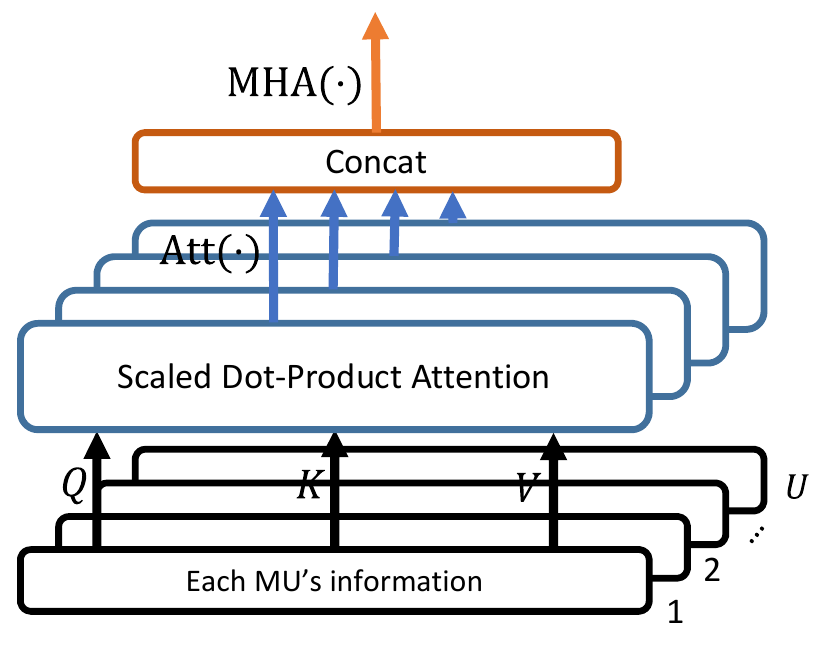}
                \caption{Structure of a Multi-Head Attention (MHA) module.}
                \label{StructureMHA}
        \end{subfigure}%
        \begin{subfigure}[t]{0.6\textwidth}
                \centering
                \captionsetup{justification=centering,singlelinecheck=false}
                \includegraphics[width=\linewidth]{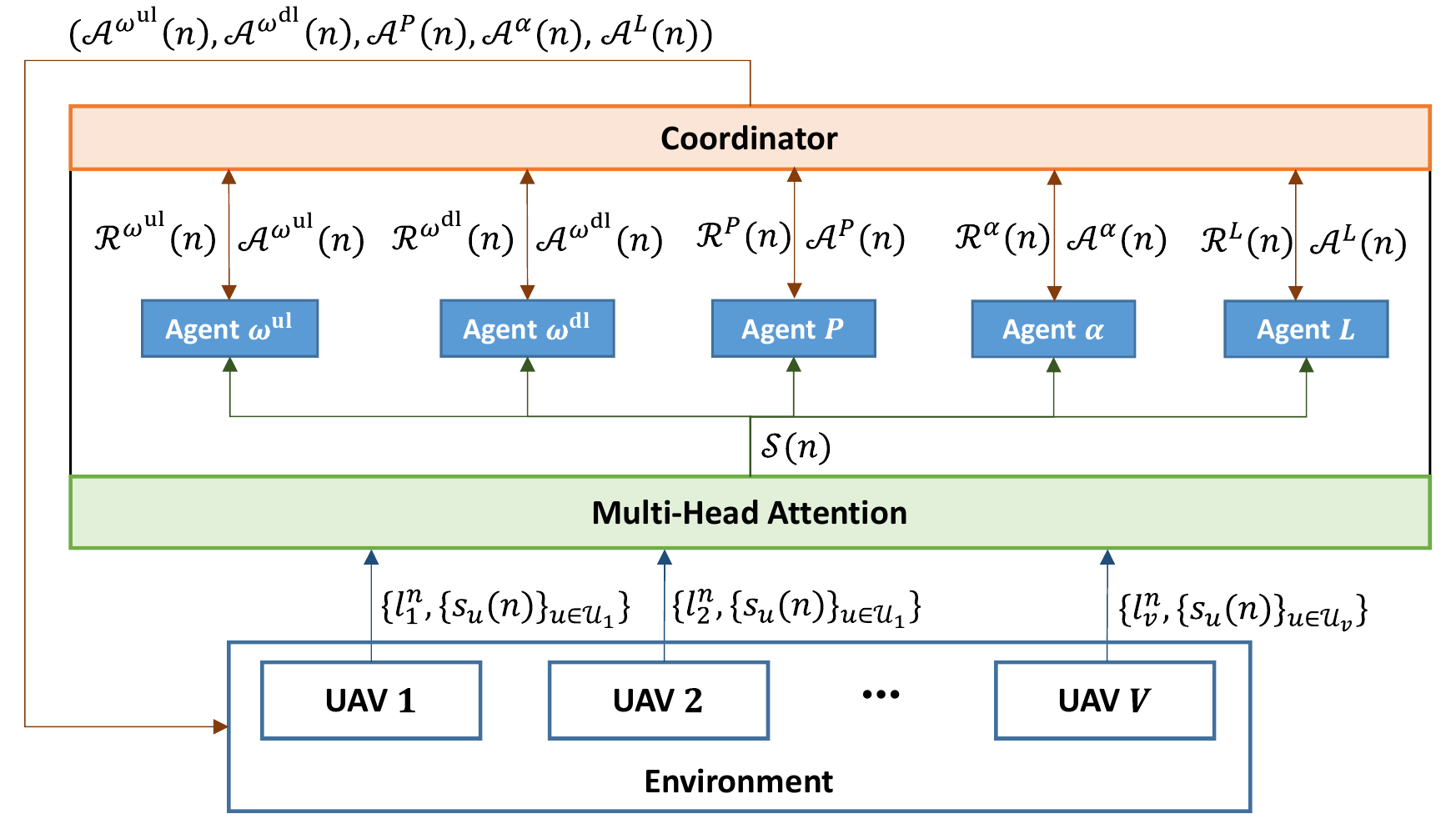}
                \caption{Structure of resources-based multi-agent proximal policy optimization (RMAPPO) with an attention mechanism.}
                \label{RMAPPO}
        \end{subfigure}%
        \caption{Illustration of the proposed RMAPPO}\label{mixed1}
\end{figure*}

\subsubsection{Embedded Multi-Head Attention (MHA) Mechanism}
In our proposed RMAPPO, we consider the ability to respond to situations in which the number of connected MUs dynamically changes. Since the input size of the general network model is fixed, we cannot effectively respond to the changing MUs information we want. Therefore, this paper proposes a learning network model regardless of the number of connected MUs by adding attention in front of the input layer. We use a technique called Multi-Head Attention (MHA) \cite{vaswani2017attention}. MHA was initially proposed as a model for natural language processing but has recently become the most popular model widely applied to deep learning-based fields such as vision and RL and natural language processing. \textcolor{black}{MHA module consists of parallel scaled dot-product attention equal to the maximum number of users that can be connected to one MEC-UAV, as shown in Fig.~\ref{StructureMHA}. Scaled dot-product attention is calculated by entering Query: $Q$, Key: $K$, and Value: $V$ of as follows:}

\textcolor{black}{
\begin{equation}
Att(Q_{u}, K_{u}, V_{u})= \textrm{softmax}({(QK^{T}) \over \sqrt{d_{u}}})V,\label{Att}
\end{equation}
where $d_{u}$ is an input's dimension of each MU $u$. In our system model, we set the input of MHA to each MU's location and task requirement. Therefore, each MU's location and requirements of task are used as Query: $Q$, Key: $K$, and Value: $V$. Through scaled dot-product attention, inputs of each MU are normalized to a fixed-sized vector, which is then combined through a $\mathrm{Concat}$ function as follows:
\begin{equation}
\textrm{MHA}(s_{u})_{u\in \mathcal{U}}= \mathrm{Concat} (Att_{1}, \dots, Att_{U}, \dots, Att_{\textrm{max}}), \label{MHA}
\end{equation}
and,
\begin{equation}
s_{u}(n) = \{\Psi_{u}(n), l^{n}_{u}\} \label{eq1},
\end{equation}
where $\Psi_{u}(n)$ represents the required information of the task before offloading at each MU $u$, and $l^{n}_{u}$ indicates the position of each MU $u$. For a nonactive scaled dot-product attention, the input is zero. As a result of the MHA module, we can optimize through the same model even if the number of MUs connected to the MEC-UAV changes. This can be used practically on the basis of the fact that the number of MUs connected to MEC-UAV can change.}

\subsubsection{Structure of Resources-based Multi-Agent Proximal Policy Optimization (RMAPPO)}
As shown in Fig.~\ref{RMAPPO}, the proposed RMAPPO in this paper is each optimization variable ($\omega, P, \alpha, L$) PPO agents. Each agent learns simultaneously in the same environment. To correspond to a different number of MUs information, the observed information is encoded into the information of a fixed size through the MHA module. Moreover, each agent derives an action to determine the reward. In this case, the MARL structure with the same reward has limitations because the constraints corresponding to each agent are different. Therefore, in this paper, the coordinator can provide different rewards according to the constraints of each agent. Next, we introduce the Markov Decision Process (MDP) of the agents used for learning.

The RMAPPO formulation for optimizing MEC-UAV network deployment is introduced based on MDP in this subsection. Initially, each agent gets the following observation based on MHA from the network:
\begin{equation}
\mathcal{S}(n) =  \{ {l}^{n}_{v}, \mathrm{MHA}(s_{u}(n))_{u\in U_{v}} \}_{u\in \mathcal{V}} \label{eq1},
\end{equation}
where $\mathrm{MHA}(\cdot)$ is the multi-head attention module, and $ {l}^{n}_{v}$ denotes MEC-UAV $v$'s position. In our proposed framework, we consider each network resource as an agent. Thus, after getting these observations, each agent takes actions according to their policy distribution, which can be defined as follows:
\begin{equation}
\mathcal{A}^{\omega^\textrm{ul}}_{v}(n)=\{\omega^{\textrm{ul}}_{u,v}(n)\}_{u\in U_{v}},
\end{equation}
\begin{equation}
\mathcal{A}^{\omega^\textrm{dl}}_{v}(n)=\{\omega^{\textrm{dl}}_{u,v}(n)\}_{u\in U_{v}},
\end{equation}
\begin{equation}
\mathcal{A}^{p}_{v}(n)=\{p_{u,v}(n)\}_{u\in U_{v}},
\end{equation}
\begin{equation}
\mathcal{A}^{L}_{v}(n)=\{\Delta{x}^{n+1}_{v}, \Delta{y}^{n+1}_{v}\},
\end{equation}
\begin{equation}
\mathcal{A}^{\alpha}_{v}(n)=\{\alpha_u(n)\}_{u\in U_{v}},
\end{equation}
where $\mathcal{A}^{\omega^\textrm{ul}}_{v}(n)$ and $\mathcal{A}^{\omega^\textrm{dl}}_{v}(n)$ denote the bandwidth allocation agent and their respective actions,  $\mathcal{A}^{p}_{v}(n)$ denotes the transmit power allocation agent and their respective actions, $\mathcal{A}^{L}_{v}(n)$ denotes the trajectory design agent and their respective actions where $\Delta{x}^{n+1}_{v}$ and $\Delta{y}^{n+1}_{v}$ are the distance traveled of x-axis and y-axis, respectably., and $\mathcal{A}^{\alpha}_{v}(n)$ denotes the amount of offloading task data control agent and their respective actions. Following the construction of the observations and action spaces, we must define a reward function that assures the goal of our optimization issue by satisfying each constraint. Therefore, in our proposed learning, the coordinator can transfer the different rewards based on each constraint to each agent. As a result, we can establish our reward function for each agent, which can be defined as follows:
\begin{equation}
\mathcal{R}^{\omega^\textrm{ul}}_{v}(n)= \Delta{u_{v}(n)} + \zeta^{\textrm{ul}} \cdot F^{v,\textrm{ul}}_{n}(f^{v, \textrm{ul}}_{n}) \cdot f^{v, \textrm{ul}}_{n},
\end{equation}
\begin{equation}
\mathcal{R}^{\omega^\textrm{dl}}_{v}(n)= \Delta{u_{v}(n)} + \zeta^{\textrm{dl}} \cdot F^{v,\textrm{dl}}_{n}(f^{v,\textrm{dl}}_{n}) \cdot f^{v,\textrm{dl}}_{n},
\end{equation}
\begin{equation}
\mathcal{R}^{p}_{v}(n)= \Delta{u_{v}(n)} + \nu \cdot G^{v}_{n}(g^{v}_{n}) \cdot g^{v}_{n},
\end{equation}
\begin{equation}
\mathcal{R}^{\alpha}_{v}(n)= \Delta{u_{v}(n)},
\end{equation}
\begin{equation}
\mathcal{R}^{L}_{v}(n)= \Delta{u_{v}(n)} - \xi \cdot h^{v}_{n},
\end{equation}
where $u_{v}(n)= \eta E_v(n) + (1-\eta)\sum_{u=1}^{U_v}  T_{u,v}(n)$ is the utility for the objective in the proposed problem, $\Delta{u_{v}(n)} = u_{v}(t)- u_{v}(t-1)$ is the sum of the utility change. $\zeta^{\textrm{ul}}$, $\zeta^{\textrm{dl}}$, $\nu$, and $\xi$ are penalty factors. $F^{v, \textrm{ul}}_{n}(f^{v, \textrm{ul}}_{n})$ is an index function that $F^{v, \textrm{ul}}_{n}(f^{v, \textrm{ul}}_{n})=0$ if $f^{v, \textrm{ul}}_{n} < 0$, and $F^{v, \textrm{ul}}_{n}(f^{v, \textrm{ul}}_{n})=1$ if $f^{v, \textrm{ul}}_{n} \geq 0$ where $f^{v, \textrm{ul}}_{n} = 1 - \sum_{u=1}^{U_v} \omega^{v, \textrm{ul}}_{u}(n)$.
$F^{v, \textrm{dl}}_{n}(f^{v, \textrm{dl}}_{n})$ is also an index function that $F^{v, \textrm{dl}}_{n}(f^{v, \textrm{dl}}_{n})=0$ if $f^{v, \textrm{dl}}_{n} < 0$, and $F^{v, \textrm{dl}}_{n}(f^{v, \textrm{dl}}_{n})=1$ if $f^{v, \textrm{dl}}_{n} \geq 0$ where $f^{v, \textrm{dl}}_{n} = 1 - \sum_{u=1}^{U_v} \omega^{v, \textrm{dl}}_{u}(n)$. Similarly, $G^{v}_{n}(g^{v}_{n})$ is an index function where $g^{v}_{n} = P_{max} - \sum_{u=1}^{U_v} \omega_{u,v}^{\textrm{dl}} B_v^{\textrm{dl}}p_{u,v}(n)$. $h^{v}_{n} \in \{ 0,1 \}$ is an indicator for constraint (\ref{C14}) that $h^{v}_{n} = 1$ if $\exists{i\in \mathcal{V}}  : \lVert {l}^{i}_{v} - {l}^{j}_{v} \rVert < L^{\mathbf{min}}$. After constructing the MDP based on our proposed optimization problem, we can obtain the optimal values of decision variables by running the proposed RMAPPO algorithm, which is defined in Algorithm \ref{alg:RMAPPO}.
\begin{algorithm}[t]
	\caption{\strut Learning Process for Resources-based Multi-Agent Proximal Policy Optimization (RMAPPO)} 
	\label{alg:RMAPPO}
	\begin{algorithmic}[1]
	    \STATE{\textbf{Initialize:} the initial network $\pi^{\omega^{\textrm{ul}}}_{0}$ and $\pi^{\omega^{\textrm{dl}}}_{0}$ for agent of subchannel, network $\pi^{p}_{0}$ for agent of power, network $\pi^{\alpha}_{0}$ for agent of offloading, and network $\pi^{L}_{0}$ for agent of trajectory}
	    \STATE{\textbf{Obtain:} optimal computation resources from \textbf{P1.1} solution, i.e., $\boldsymbol{c}^{*\mathrm{in}}$ and $\boldsymbol{c}^{*\mathrm{mec}}$}
		\FOR{episode$=1,2,...,E$}
		\STATE{Initialize randomly each MU's position and association.}
		\FOR{time slot$=1,2,...,N$}
		\STATE{Upload observation $s_{v}(n)$ of each UAV-BS $v$ to Airship}
		\STATE{Encoding $s_{v}(n)$ to $\mathcal{S}_{v} = \textrm{MHA}(s_{v}(n)_{v\in\mathcal{V}})$}
		\STATE{Run policy $\mathcal{A}^{\omega^{\textrm{ul}}}_{v}\sim\pi^{\omega^{\textrm{ul}}}_{\theta_{\textrm{old}}}$ $\mathcal{A}^{\omega^{\textrm{dl}}}_{v}\sim\pi^{\omega^{\textrm{dl}}}_{\theta_{\textrm{old}}}$, $\mathcal{A}^{p}_{v}\sim\pi^{p}_{\theta_{\textrm{old}}}$, $\mathcal{A}^{\alpha}_{v}\sim\pi^{\alpha}_{\theta_{\textrm{old}}}$, and $\mathcal{A}^{L}_{v}\sim\pi^{L}_{\theta_{\textrm{old}}}$.}
		\STATE{Compute each reward $\mathcal{R}^{\omega^{\textrm{ul}}}_{v}$, $\mathcal{R}^{\omega^{\textrm{dl}}}_{v}$, $\mathcal{R}^{p}_{v}$, $\mathcal{R}^{\alpha}_{v}$, and $\mathcal{R}^{L}_{v}$ for UAV-BS $v$}
		\STATE{Save $(\mathcal{S}_{v}(n),\mathcal{A}_{v}(n),\mathcal{R}_{v}(n),\mathcal{S}_{v}(n+1))$ in memory of each agent}
		\ENDFOR
		\STATE{Compute advantage estimates $\left\langle \hat{A}^{\omega^{\textrm{ul}}}_{1},...,\hat{A}^{\omega^{\textrm{ul}}}_{N} \right\rangle$, $\left\langle \hat{A}^{\omega^{\textrm{dl}}}_{1},...,\hat{A}^{\omega^{\textrm{dl}}}_{N} \right\rangle$, $\left\langle \hat{A}^{p}_{1},...,\hat{A}^{p}_{N} \right\rangle$, $\left\langle \hat{A}^{\alpha}_{1},...,\hat{A}^{\alpha}_{N} \right\rangle$, and $\left\langle \hat{A}^{\textbf{L}}_{1},...,\hat{A}^{\textbf{L}}_{N} \right\rangle$}
		\STATE{Optimize surrogate $L^{\textrm{PPO}}$ wrt $\theta^{\omega^{\textrm{ul}}}$, $\theta^{\omega^{\textrm{dl}}}$, $\theta^{p}$, $\theta^{\alpha}$, and $\theta^{L}$, with minibatch from memory}
		\STATE{$\theta^{\omega^{\textrm{ul}}}_{\textrm{old}} \leftarrow \theta^{\omega^{\textrm{ul}}}$, $\theta^{\omega^{\textrm{dl}}}_{\textrm{old}} \leftarrow \theta^{\omega^{\textrm{dl}}}$, $\theta^{p}_{\textrm{old}} \leftarrow \theta^{p}$, $\theta^{\alpha}_{\textrm{old}} \leftarrow \theta^{\alpha}$, and $\theta^{L}_{\textrm{old}} \leftarrow \theta^{L}$ }
		\ENDFOR
		\STATE{\textbf{Output:} Optimal networks $\pi^{\omega^{\textrm{ul}}}_{\theta_{\textrm{opt}}}$, $\pi^{\omega^{\textrm{dl}}}_{\theta_{\textrm{opt}}}$, $\pi^{p}_{\theta_{\textrm{opt}}}$, $\pi^{\alpha}_{\theta_{\textrm{opt}}}$, and $\pi^{L}_{\theta_{\textrm{opt}}}$}
	\end{algorithmic}
\end{algorithm}
\begin{table}[t]
	\caption{\textbf{Parameters for Simulation}}
	\setlength{\tabcolsep}{2.6pt}
	{\footnotesize
		\renewcommand{\arraystretch}{1.2}
		\begin{tabular}{|p{60pt}| p{120pt}| p{55pt}|}
			\hline
			\textbf{Parameters} & \textbf{Description}  & \textbf{Value}	\\
			\hline
			$\sigma^{2}$&Additive white Gaussian noise power&-175 dBm/Hz\\
			\hline
			$p_0$&Uplink transmitting power of each users& 0.5 W\\
			\hline
			$P^{\mathbf{max}}$&Total transmitting power for downlink of each UAVs& 5 W\\
			\hline
			$a$&Absorption coefficient&0.005\\
			\hline
			$B$&Total bandwidth&0.2 THz\\
			\hline
			$R^{\mathbf{min}}$&Minimum achievable rate&0.05 Tbps\\
			\hline
			$h_0$&Channel gain at ref.&-40 dBm\\
			\hline
			$T$&Number of time slots in each episode&100\\
		\end{tabular}
		\begin{tabular}{|p{186pt}| p{55pt}|}
			\hline
			Batch size& 256\\
			\hline
			Learning rate& 0.0003\\
			\hline
			Max steps& 200000\\
			\hline
			Number of units for each hidden layer& 256\\
			\hline
			Number of hidden layer for $\textrm{MAPPO}_{\textrm{RS}}$&2\\
			\hline
			Number of hidden layer for $\textrm{MAPPO}_{\textrm{UAV}}$&3\\
			\hline
		\end{tabular}
	}
	\label{tab1}
\end{table}
\section{Performance Evaluation}
\label{simulation_results}
\subsection{Simulation Setup \& Benchmarks}
We consider $3$ MEC-UAVs in a region of $500~$m$^2$, which provide services to the $50$ MUs in our simulation setup. It is assumed that MEC-Enabled UAVs can initially move at a fixed altitude of $h_0=50~$m. Moreover, the learning network model's parameters and main network parameters' values are given in Table \ref{tab1}. Finally, to evaluate the performance of the proposed algorithm, we use the following four benchmark algorithms as follows:
\begin{itemize}
    \item \emph{RMAPPO (proposed)}: The proposed resources-based MAPPO algorithm, in which each optimization variable is thought of as an agent that learns its variable optimization.
    \item \emph{GMAPPO}: The General MAPPO algorithm, in which one MEC-Enabled UAV acts as an agent and learns resource allocation and trajectory optimization simultaneously.
    \item $\emph{Fairness}_{\emph{All}}$: The algorithm takes into account a fair all-resource allocation to each MU with a trajectory that leads directly to centroids.
    \item $\emph{Fairness}_{\emph{W}}$: This algorithm takes into account a fair sub-band allocation for each MU. However, it optimizes variables except for sub-band allocation using the proposed algorithm.
    \item $\emph{Fairness}_{\emph{P}}$: This method takes into account a fair power allocation to each MU. However, it optimizes factors other than power allocation using the suggested approach.
\end{itemize}
\subsection{Simulation Results \& Discussion}
\begin{figure}[t]
    \centering
    \includegraphics[width=0.8\columnwidth]{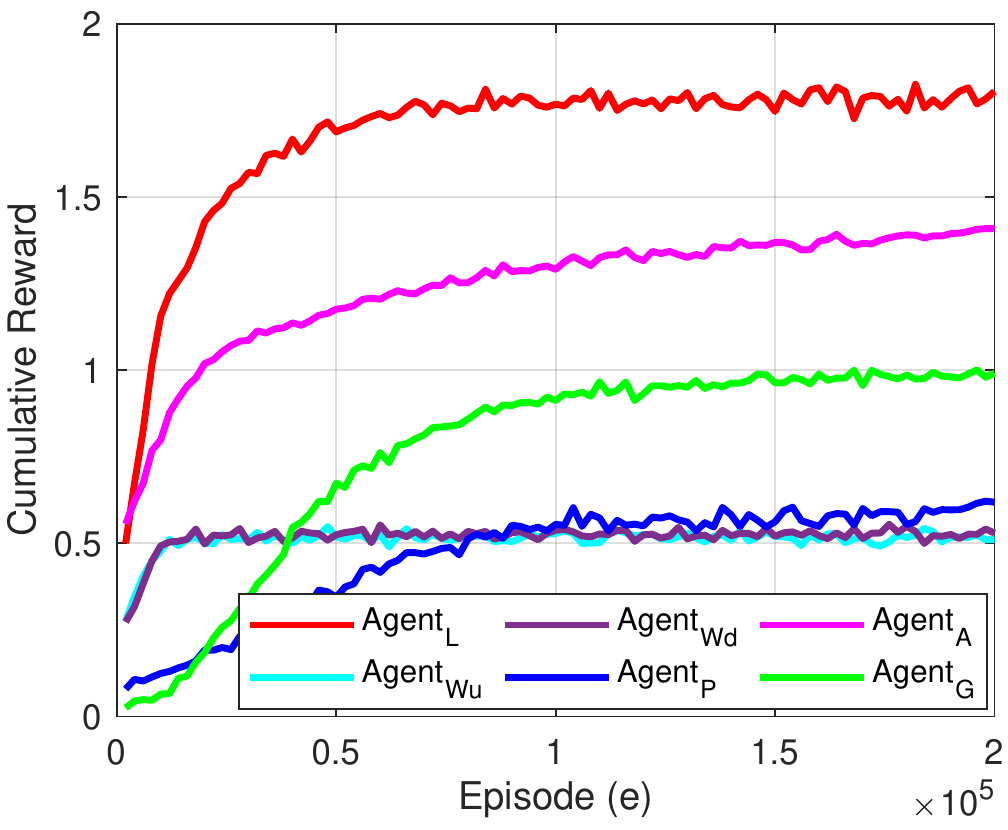}
    \caption{Cumulative rewards over the number of episode.}
    \label{simulation_result_1}
\end{figure}
\begin{figure*}[t]
    \centering
    \includegraphics[width=1.0\textwidth]{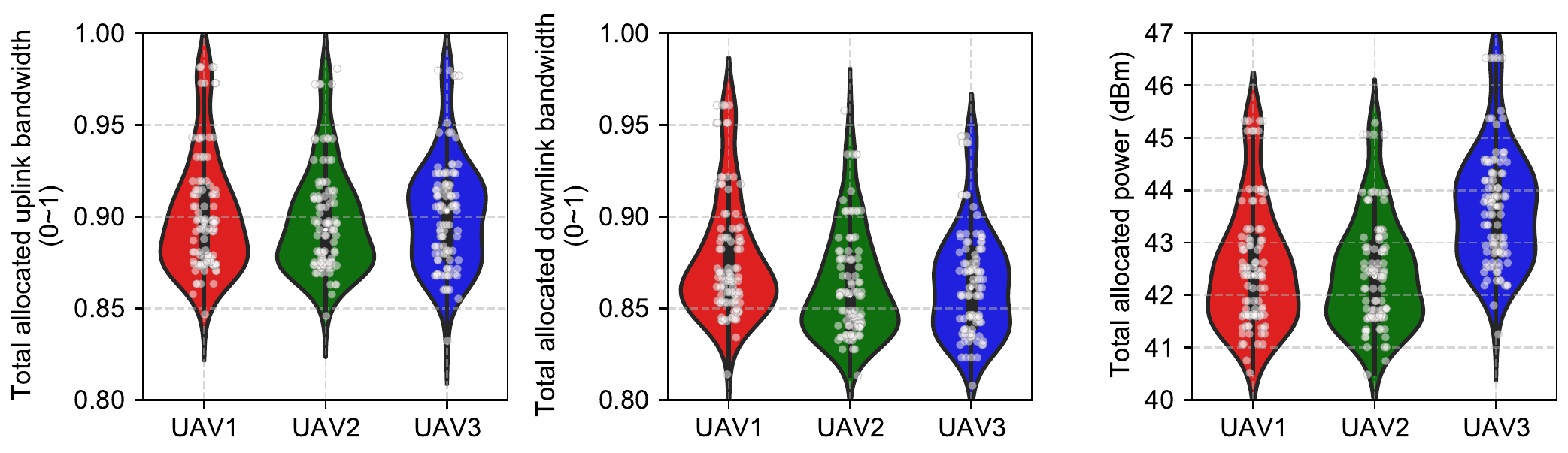}
    \caption{Distribution of the total allocated resource for each time.}
    \label{simulation_result_4}
\end{figure*}
\begin{figure}[t]
    \centering
    \includegraphics[width=0.8\columnwidth]{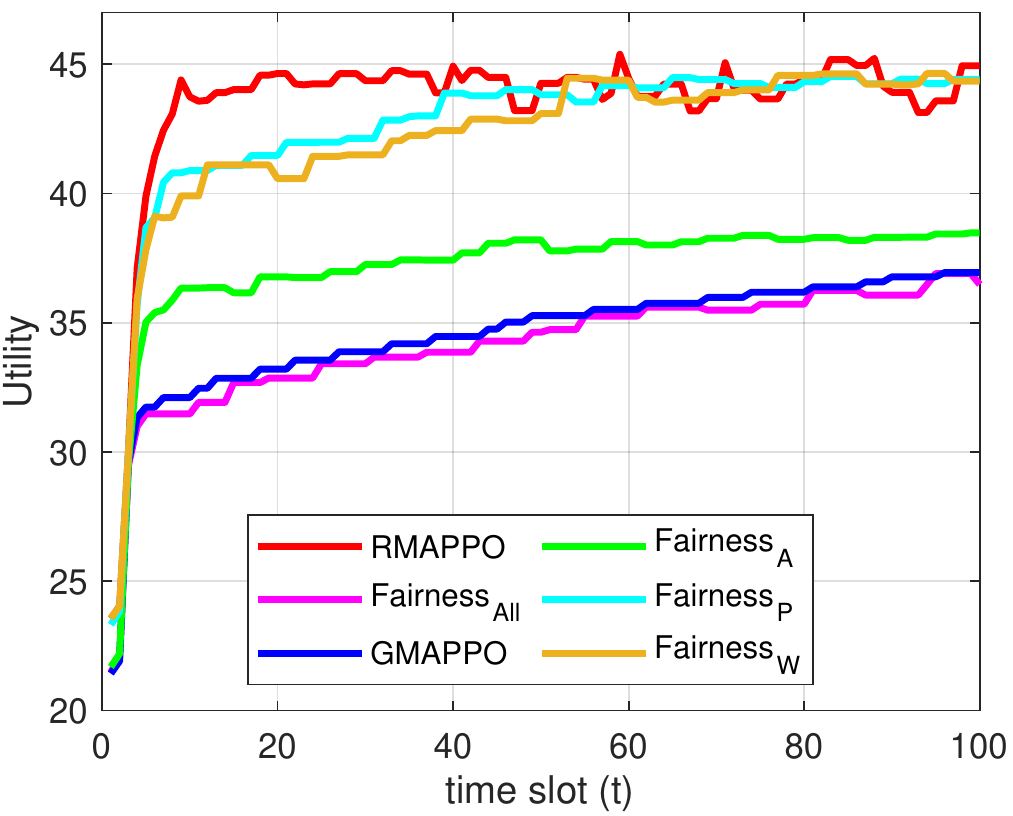}
    \caption{Comparison of utility with benchmarks schemes over the time slots.}
    \label{simulation_result_3}
\end{figure}
Fig.~\ref{simulation_result_1} depicts the cumulative reward of each learning agent as the episode progresses. With the suggested $\textrm{RMAPPO}$, $\textrm{Agent}^{*}$ is accountable for each resource variable ($\boldsymbol{\omega^{\textrm{ul}}}, \boldsymbol{\omega^{\textrm{dl}}}, \boldsymbol{p}, \boldsymbol{L}, \boldsymbol{\alpha}$). At the same time, $\textrm{Agent}_{\textrm{G}}$ learns all resource variables according to the $\textrm{GMAPPO}$. As seen in Fig.~\ref{simulation_result_1}, all agents converge around $1e+5$ episodes. It can be shown that $\textrm{Agent}_{\textrm{L}}$ has the largest final cumulative reward, allowing us to conclude that trajectory optimization has the biggest influence on utility when compared to other resource variables. Also, because the learning is not entirely optimized, $\textrm{Agent}t_{\textrm{G}}$ has a final cumulative reward of around $1$, which is smaller than $\textrm{Agent}_{\textrm{A}}$.

Fig.~\ref{simulation_result_4} shows the distribution of resources allocated over the total time slot using a violin plot. In Fig.~\ref{simulation_result_4}, the maximum allocations of uplink and downlink bandwidth are less than $1$, and the power is less than $47$ dBm. As a result, we can see that resource allocation through the proposed algorithm satisfies the constraints and is allocated successfully.

The trained model is used to depict the change in utility of each benchmark algorithm over time in the same experimental setting as depicted in Fig.~\ref{simulation_result_3}. To begin with, all benchmark algorithms exhibit graphs of increasing utility with time. This is because MEC-enabled UAVs will become more useful as they move closer to the MU. In terms of convergence speed and final utility, the proposed RMAPPO outperforms all benchmarks. Following that, $\textrm{Fairness}_{\textrm{W}}$ and $\textrm{Fairness}_{\textrm{P}}$ perform well since it is sensible to provide fair resources when the distance between the MEC-Enabled UAV and MUs grows. Following that, the fact that $\textrm{Fairness}_{\textrm{A}}$ has poor performance demonstrates that offloading is a key variable for utilities. Finally, GMAPPO performs similarly or somewhat better than $\textrm{Fairness}_{\textrm{All}}$. Because the number of hidden layers or units in the learning model was inadequate, many resource variables could not be fully optimized simultaneously, resulting in underfitting.

\begin{figure}[t]
    \centering
    \includegraphics[width=0.9\columnwidth]{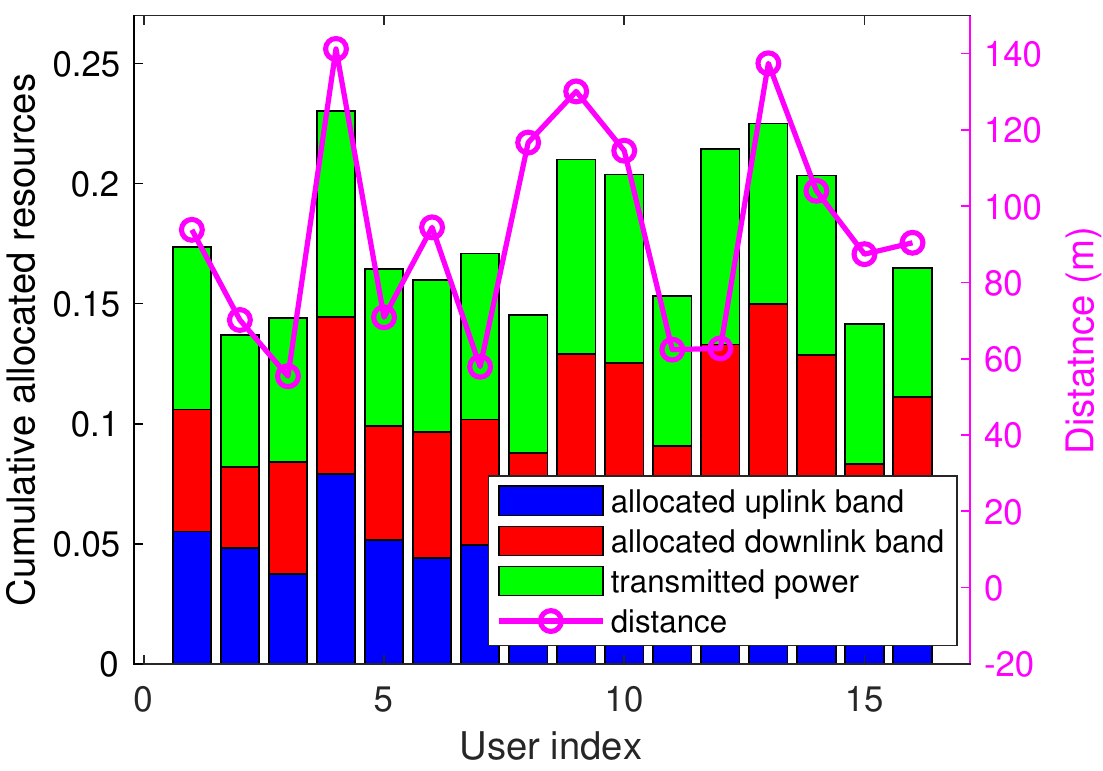}
    \caption{Resources allocation to each MU with distance.}
    \label{simulation_result_5}
\end{figure}
\begin{figure}[htb!]
    \centering
    \includegraphics[width=0.8\columnwidth]{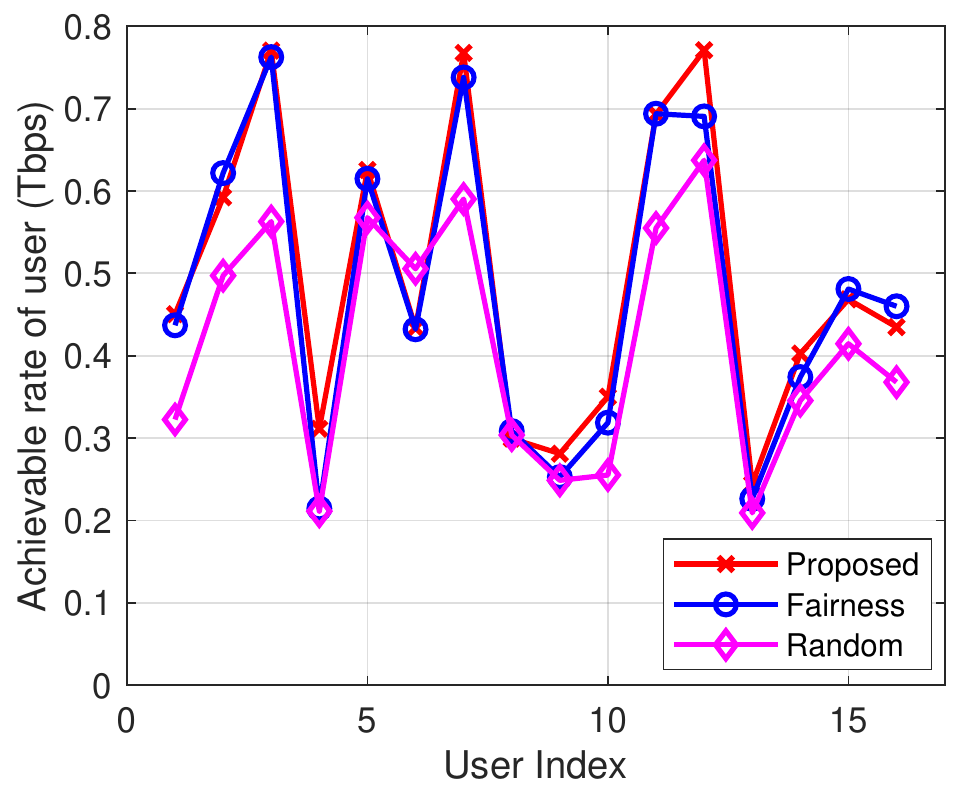}
    \caption{Achievable rates of each MU with the time slot.}
    \label{simulation_result_6}
\end{figure}

The distances and cumulative allotted resources bar graph of each MU belonging to a MEC-Enabled UAV in a one-time slot are shown in Fig.~\ref{simulation_result_5}. It can be observed in Fig.~\ref{simulation_result_5}, that the proposed RMAPPO has a propensity to award more resources to MUs who are far away. Fig.~\ref{simulation_result_6} shows a beneficial effect on this, i.e., the feasible rates of each MU in the same environment as presented in Fig.~\ref{simulation_result_5} can be validated in Fig.~\ref{simulation_result_6}. The proposed resource optimization algorithm resulted in an average $3.53\% $ and $19.74\% $ performance improvement in the achievable rate of MUs compared to the fair and random resource allocation, respectively.

\begin{figure}[t]
    \centering
    \includegraphics[width=0.9\columnwidth]{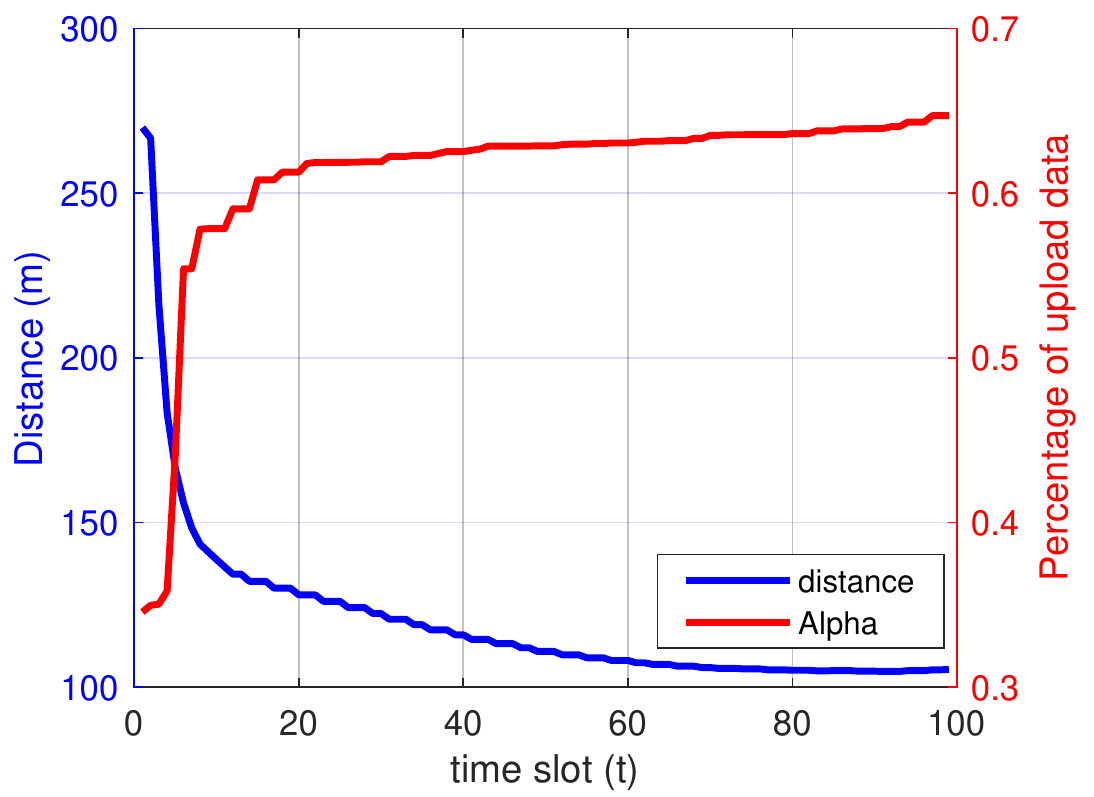}
    \caption{Comparison with alpha and distance over the time slot}
    \label{simulation_result_7}
\end{figure}

Fig.~\ref{simulation_result_7} depicts the offloading variable $\alpha$ of one MU based on time slot. The comparison graph in Fig.~\ref{simulation_result_7} indicates a trade-off relationship between distance and offloading. It can be observed that the closer the distance, the more the MU offloads the MEC-enabled UAV to boost performance.

\begin{figure}[t]
    \centering
    \includegraphics[width=0.8\columnwidth]{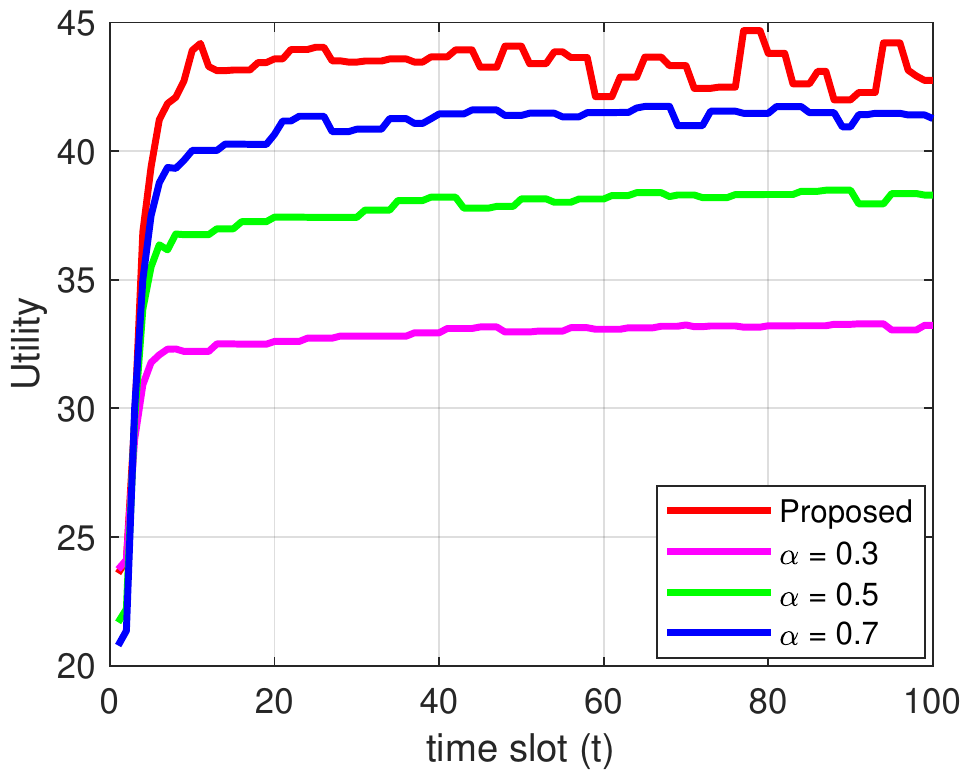}
    \caption{Utility of different alpha over the time slot }
    \label{simulation_result_8}
\end{figure}
Finally, Fig.~\ref{simulation_result_8} compares the performance of the optimized offloading variable $\alpha$ to the fixed offloading ratio ($\alpha=0.3,0.5,0.7$) using the proposed RMAPPO. As a consequence, the proposed algorithms improve average performance by $30.45\%$, $13.97\%$, and $5.22\%$ compared to the fixed offloading ratio. Moreover, by comparing performance in the early and late time slots, it is discovered that it is more effective to conduct less offloading in the early time slot when the distance between the MEC-Enabled UAV and the MU is significant, and it is more efficient to do more offloading when the distance is small.\\
\begin{figure}[t]
    \centering
    \includegraphics[width=0.8\columnwidth]{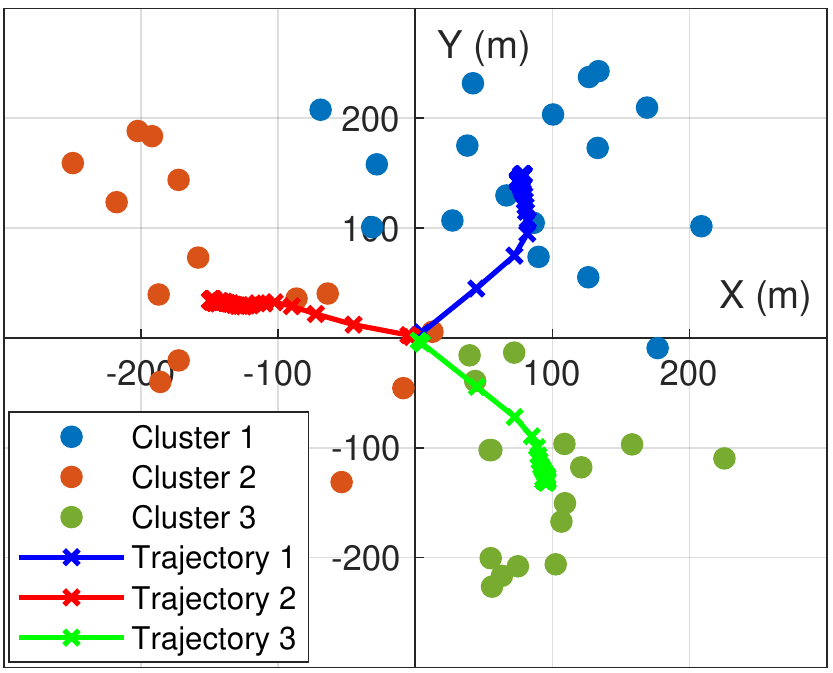}
    \caption{Deployment of MEC-enabled UAV-BSs with trained model for MUs.}
    \label{simulation_result_2}
\end{figure}
The trajectories of MEC-Enabled UAVs produced from the training model and the MU's deployments are shown in Fig.~\ref{simulation_result_2}. The generated trajectory goes closer to the centroid of each cluster, demonstrating the efficiency of our suggested methods for trajectory optimization.
\section{Conclusion}
\label{conclusion}
This paper investigated the joint resource allocation and trajectory optimization of MEC-UAVs in sub-THz networks. Then, we devised an optimization problem to minimize the sum of MEC-UAVs and MUs energy consumption and task delay under the given task information while meeting resource constraints. To deal with this issue, we divided the original problem into two subproblems. Then, we proposed an RMAPPO algorithm to deal jointly with resource allocation and trajectory optimization, which can make quick decisions in the given environment due to its low complexity. Furthermore, each agent can focus more on their actions and rewards based on the proposed RMAPPO. Furthermore, the proposed system can respond effectively to a wide range of observation sizes by taking into account the attention mechanism. As presented in the simulation results, our proposed algorithm outperforms the other benchmarks and yields a network utility of $2.22\%$, $15.55\%$, and $17.77\%$ better than the benchmarks.
\ifCLASSOPTIONcaptionsoff
  \newpage
\fi
\bibliographystyle{IEEEtran}
\bibliography{ref}

\begin{thebibliography}{10}
\providecommand{\url}[1]{#1}
\csname url@samestyle\endcsname
\providecommand{\newblock}{\relax}
\providecommand{\bibinfo}[2]{#2}
\providecommand{\BIBentrySTDinterwordspacing}{\spaceskip=0pt\relax}
\providecommand{\BIBentryALTinterwordstretchfactor}{4}
\providecommand{\BIBentryALTinterwordspacing}{\spaceskip=\fontdimen2\font plus
\BIBentryALTinterwordstretchfactor\fontdimen3\font minus
  \fontdimen4\font\relax}
\providecommand{\BIBforeignlanguage}[2]{{%
\expandafter\ifx\csname l@#1\endcsname\relax
\typeout{** WARNING: IEEEtran.bst: No hyphenation pattern has been}%
\typeout{** loaded for the language `#1'. Using the pattern for}%
\typeout{** the default language instead.}%
\else
\language=\csname l@#1\endcsname
\fi
#2}}
\providecommand{\BIBdecl}{\relax}
\BIBdecl

\bibitem{Salman_globecom}
S.~S. Hassan, Y.~K. Tun, W.~Saad, Z.~Han, and C.~S. Hong, ``Blue data
  computation maximization in {6G} space-air-sea non-terrestrial networks,'' in
  \emph{the proc. of IEEE Global Communications Conference (GLOBECOM)}, Madrid,
  Spain, Dec. 2021.

\bibitem{own_NOMS}
Y.~M. Park, S.~S. Hassan, Y.~K. Tun, Z.~Han, and C.~S. Hong, ``Joint resources
  and phase-shift optimization of {MEC}-enabled {UAV} in {IRS}-assisted {6G}
  {TH}z networks,'' in \emph{the proc. of IEEE/IFIP Network Operations and
  Management Symposium (NOMS)}, Budapest, Hungary, Apr. 2022.

\bibitem{Salman_Iot}
S.~S. Hassan, D.~H. Kim, Y.~K. Tun, N.~H. Tran, W.~Saad, and C.~S. Hong,
  ``Seamless and energy efficient maritime coverage in coordinated {6G}
  space-air-sea non-terrestrial networks,'' \emph{arXiv preprint
  arXiv:2201.08605}, Jan. 2022.

\bibitem{traffic_forecast}
H.~Shin, J.~Jung, and Y.~Koo, ``{Forecasting the video data traffic of {5G}
  services in {S}outh {K}orea},'' \emph{Technological Forecasting and Social
  Change}, vol. 153, p. 119948, Apr. 2020.

\bibitem{THz_survey}
H.~Elayan, O.~Amin, R.~M. Shubair, and M.-S. Alouini, ``Terahertz
  communication: {T}he opportunities of wireless technology beyond {5G},'' in
  \emph{the proc. of International Conference on Advanced Communication
  Technologies and Networking (CommNet)}, Marrakech, Morocco, May 2018.

\bibitem{technologies7020043}
J.~F. O’Hara, S.~Ekin, W.~Choi, and I.~Song, ``A perspective on {T}erahertz
  next-generation wireless communications,'' \emph{Technologies}, vol.~7,
  no.~2, Jun. 2019.

\bibitem{THz_pathloss}
N.~Elburki, S.~Ben~Amor, and S.~Affes, ``Evaluation of path-loss models for
  {TH}z propagation in indoor environments,'' in \emph{the proc. of IEEE
  Canadian Conference on Electrical and Computer Engineering (CCECE)}, London,
  ON, Canada, Sep. 2020.

\bibitem{THz_system}
C.~Lin and G.~Y. Li, ``Indoor terahertz communications: {H}ow many antenna
  arrays are needed?'' \emph{IEEE Transactions on Wireless Communications},
  vol.~14, no.~6, pp. 3097--3107, Feb. 2015.

\bibitem{mimo_antenna}
X.~Gao, L.~Dai, S.~Han, C.-L. I, and R.~W. Heath, ``Energy-efficient hybrid
  analog and digital precoding for mm{W}ave {MIMO} systems with large antenna
  arrays,'' \emph{IEEE Journal on Selected Areas in Communications}, vol.~34,
  no.~4, pp. 998--1009, Mar. 2016.

\bibitem{THz_link_budget}
H.~Elayan, R.~M. Shubair, J.~M. Jornet, and P.~Johari, ``Terahertz channel
  model and link budget analysis for intrabody nanoscale communication,''
  \emph{IEEE Transactions on NanoBioscience}, vol.~16, no.~6, pp. 491--503,
  Jun. 2017.

\bibitem{power_allocation_THz}
O.~Elkharbotly, E.~Maher, A.~El-Mahdy, and F.~Dressler, ``Optimal power
  allocation in cooperative {MIMO-NOMA} with {FD/HD} relaying in {TH}z
  communications,'' in \emph{the proc. of 9th IFIP International Conference on
  Performance Evaluation and Modeling in Wireless Networks (PEMWN)}, Berlin,
  Germany, Dec. 2020.

\bibitem{THz_various_altitudes}
A.~Saeed, O.~Gurbuz, and M.~A. Akkas, ``Terahertz communications at various
  atmospheric altitudes,'' \emph{Physical Communication}, vol.~41, p. 101113,
  Aug. 2020.

\bibitem{THz_absorbtion}
D.~Oluseun, S.~Thomas, O.~Idowu-Bismark, P.~Nzerem, and I.~Muhammad,
  ``Absorption, diffraction and free space path losses modeling for the
  {T}erahertz band,'' \emph{International Journal of Engineering and
  Manufacturing}, vol.~10, pp. 54--65, Feb. 2020.

\bibitem{mobility_UAV_effect}
Z.~Guan and T.~Kulkarni, ``On the effects of mobility uncertainties on wireless
  communications between flying drones in the mm{W}ave/{TH}z bands,'' in
  \emph{the proc. of IEEE Conference on Computer Communications Workshops
  (INFOCOM WKSHPS)}, Paris, France, Apr. 2019.

\bibitem{errors_THz_UAV_positioning}
R.~Mendrzik, D.~Cabric, and G.~Bauch, ``Error bounds for {T}erahertz {MIMO}
  positioning of swarm {UAV}s for distributed sensing,'' in \emph{the proc. of
  IEEE International Conference on Communications Workshops (ICC Workshops)},
  Kansas City, MO, May 2018.

\bibitem{Salman_ICC}
S.~S. Hassan, Y.~M. Park, Y.~Kyaw~Tun, W.~Saad, Z.~Han, and C.~S. Hong,
  ``{3TO}: {TH}z-enabled throughput and trajectory optimization of {UAV}s in
  {6G} networks by proximal policy optimization deep reinforcement learning,''
  in \emph{the proc. of IEEE International Conference on Communications (ICC)},
  Seoul, South Korea, May. 2022.

\bibitem{full-duplex1}
A.~Tang and X.~Wang, ``A-duplex: Medium access control for efficient
  coexistence between full-duplex and half-duplex communications,'' \emph{IEEE
  Transactions on Wireless Communications}, vol.~14, no.~10, pp. 5871--5885,
  Jun. 2015.

\bibitem{full-duplex2}
F.~Passerini and A.~M. Tonello, ``Analog full-duplex amplify-and-forward relay
  for power line communication networks,'' \emph{IEEE Communications Letters},
  vol.~23, no.~4, pp. 676--679, Feb. 2019.

\bibitem{Full-duplex3}
T.~P. Do and Y.~H. Kim, ``Resource allocation for a full-duplex
  wireless-powered communication network with imperfect self-interference
  cancelation,'' \emph{IEEE Communications Letters}, vol.~20, no.~12, pp.
  2482--2485, Sep. 2016.

\bibitem{Full-duplex4}
D.~Wang, M.~Wang, P.~Zhu, J.~Li, J.~Wang, and X.~You, ``Performance of
  network-assisted full-duplex for cell-free massive {MIMO},'' \emph{IEEE
  Transactions on Communications}, vol.~68, no.~3, pp. 1464--1478, Dec. 2020.

\bibitem{Full-duplex5}
M.~Liu, Y.~Mao, and S.~Leng, ``Cooperative fog-cloud computing enhanced by
  full-duplex communications,'' \emph{IEEE Communications Letters}, vol.~22,
  no.~10, pp. 2044--2047, Aug. 2018.

\bibitem{Full-duplex6}
J.~Li, X.~Li, A.~Wang, and N.~Ye, ``Beamspace {MIMO-NOMA} for millimeter-wave
  broadcasting via full-duplex {D2D} communications,'' \emph{IEEE Transactions
  on Broadcasting}, vol.~66, no.~2, pp. 545--554, Mar. 2020.

\bibitem{THz-band1}
X.~Zhang, J.~Wang, and H.~V. Poor, ``Joint resource allocation optimization
  over energy harvesting based {6G} {TH}z-band big-data-driven nano-networks,''
  in \emph{the proc. of IEEE Global Communications Conference (GLOBECOM)},
  Taipei, Taiwan, Dec. 2020.

\bibitem{THz-band2}
V.~Petrov, J.~Kokkoniemi, D.~Moltchanov, J.~Lehtomäki, M.~Juntti, and
  Y.~Koucheryavy, ``The impact of interference from the side lanes on
  mm{W}ave/{TH}z band {V2V} communication systems with directional antennas,''
  \emph{IEEE Transactions on Vehicular Technology}, vol.~67, no.~6, pp.
  5028--5041, Jan. 2018.

\bibitem{THz-band3}
N.~Khalid and O.~B. Akan, ``Experimental throughput analysis of low-{TH}z
  {MIMO} communication channel in {5G} wireless networks,'' \emph{IEEE Wireless
  Communications Letters}, vol.~5, no.~6, pp. 616--619, Sep. 2016.

\bibitem{THz-band4}
N.~A. Abbasi, A.~Hariharan, A.~M. Nair, A.~S. Almaiman, F.~B. Rottenberg, A.~E.
  Willner, and A.~F. Molisch, ``Double directional channel measurements for
  {TH}z communications in an urban environment,'' in \emph{the proc. of IEEE
  International Conference on Communications (ICC)}, Dublin, Ireland, Jun.
  2020.

\bibitem{UAV_1}
J.~Baek, S.~I. Han, and Y.~Han, ``Energy-efficient uav routing for wireless
  sensor networks,'' \emph{IEEE Transactions on Vehicular Technology}, vol.~69,
  no.~2, pp. 1741--1750, Dec. 2020.

\bibitem{UAV_2}
Z.~Wang, W.~Xu, D.~Yang, and J.~Lin, ``Joint trajectory optimization and user
  scheduling for rotary-wing {UAV}-enabled wireless powered communication
  networks,'' \emph{IEEE Access}, vol.~7, pp. 181\,369--181\,380, Dec. 2019.

\bibitem{UAV_3}
Z.~Wei, J.~Zhu, Z.~Guo, and F.~Ning, ``The performance analysis of spectrum
  sharing between {UAV} enabled wireless mesh networks and ground networks,''
  \emph{IEEE Sensors Journal}, vol.~21, no.~5, pp. 7034--7045, Nov. 2021.

\bibitem{UAV_4}
H.-T. Ye, X.~Kang, J.~Joung, and Y.-C. Liang, ``Optimization for full-duplex
  rotary-wing {UAV}-enabled wireless-powered {I}o{T} networks,'' \emph{IEEE
  Transactions on Wireless Communications}, vol.~19, no.~7, pp. 5057--5072,
  Apr. 2020.

\bibitem{UAV_5}
Y.~K. Tun, T.~N. Dang, K.~Kim, M.~Alsenwi, W.~Saad, and C.~S. Hong,
  ``Collaboration in the sky: {A} distributed framework for task offloading and
  resource allocation in multi-access edge computing,'' \emph{IEEE Internet of
  Things Journal}, pp. 1--1, July 2022.

\bibitem{pmlr-v139-liu21j}
W.~Mao, L.~Yang, K.~Zhang, and T.~Basar, ``On improving model-free algorithms
  for decentralized multi-agent reinforcement learning,'' in
  \emph{International Conference on Machine Learning}, Maryland, USA, Jul.
  2022.

\bibitem{9509579}
Y.~Ye, Y.~Tang, H.~Wang, X.-P. Zhang, and G.~Strbac, ``A scalable
  privacy-preserving multi-agent deep reinforcement learning approach for
  large-scale peer-to-peer transactive energy trading,'' \emph{IEEE
  Transactions on Smart Grid}, vol.~12, no.~6, pp. 5185--5200, Aug. 2021.

\bibitem{liu2021cooperative}
I.-J. Liu, U.~Jain, R.~A. Yeh, and A.~Schwing, ``Cooperative exploration for
  multi-agent deep reinforcement learning,'' in \emph{International Conference
  on Machine Learning}, Online, Jul. 2021.

\bibitem{MADRL_UAV}
J.~Cui, Y.~Liu, and A.~Nallanathan, ``Multi-agent reinforcement learning-based
  resource allocation for {UAV} networks,'' \emph{IEEE Transactions on Wireless
  Communications}, vol.~19, no.~2, pp. 729--743, Aug. 2020.

\bibitem{xu2021joint}
L.~Xu, M.~Chen, M.~Chen, Z.~Yang, C.~Chaccour, W.~Saad, and C.~S. Hong, ``Joint
  location, bandwidth and power optimization for {TH}z-enabled {UAV}
  communications,'' \emph{IEEE Communications Letters}, vol.~25, no.~6, pp.
  1984--1988, Mar. 2021.

\bibitem{tun2020energy}
Y.~K. Tun, Y.~M. Park, N.~H. Tran, W.~Saad, S.~R. Pandey, and C.~S. Hong,
  ``Energy-efficient resource management in {UAV}-assisted mobile edge
  computing,'' \emph{IEEE Communications Letters}, vol.~25, no.~1, pp.
  249--253, Sep. 2020.

\bibitem{hu2018joint}
Q.~Hu, Y.~Cai, G.~Yu, Z.~Qin, M.~Zhao, and G.~Y. Li, ``Joint offloading and
  trajectory design for {UAV}-enabled mobile edge computing systems,''
  \emph{IEEE Internet of Things Journal}, vol.~6, no.~2, pp. 1879--1892, Oct.
  2018.

\bibitem{li2020delay}
L.~Li, M.~Wang, K.~Xue, Q.~Cheng, D.~Wang, W.~Chen, M.~Pan, and Z.~Han, ``Delay
  optimization in multi-{UAV} edge caching networks: {A} robust mean field
  game,'' \emph{IEEE Transactions on Vehicular Technology}, vol.~70, no.~1, pp.
  808--819, Dec. 2020.

\bibitem{diamond2016cvxpy}
S.~Diamond and S.~Boyd, ``{CVXPY}: {A} {P}ython-embedded modeling language for
  convex optimization,'' \emph{Journal of Machine Learning Research}, vol.~17,
  no.~83, pp. 1--5, Jan. 2016.

\bibitem{ppo}
J.~Schulman, F.~Wolski, P.~Dhariwal, A.~Radford, and O.~Klimov, ``Proximal
  policy optimization algorithms,'' \emph{arXiv preprint arXiv:1707.06347},
  2017.

\bibitem{trpo}
J.~Schulman, S.~Levine, P.~Abbeel, M.~Jordan, and P.~Moritz, ``Trust region
  policy optimization,'' in \emph{the proc. of International conference on
  machine learning (ICML)}, Lille, France, Jul. 2015.

\bibitem{vaswani2017attention}
A.~Vaswani, N.~Shazeer, N.~Parmar, J.~Uszkoreit, L.~Jones, A.~N. Gomez,
  {\L}.~Kaiser, and I.~Polosukhin, ``Attention is all you need,''
  \emph{Advances in neural information processing systems}, vol.~30, 2017.

\end{thebibliography}
\end{document}